\documentclass[sigconf]{acmart}
\def\BibTeX{{\rm B\kern-.05em{\sc i\kern-.025em b}\kern-.08emT\kern-.1667em\lower.7ex\hbox{E}\kern-.125emX}}

\renewcommand\footnotetextcopyrightpermission[1]{} 
\setcopyright{none}

\usepackage{tikz}

\usepackage[font={bf,sf,small}]{subfig}
\usepackage[font={bf,sf,small}]{caption}
\usepackage{amsmath}
\usepackage{listings}

\usepackage{mathtools}

\usepackage{enumitem}
\usepackage{booktabs}
\usepackage{multirow}
\usepackage{makecell}

\newlength{\verticalcompensationlength}
\newcounter{verticalcompensationrows}
\newcommand{\verticalcompensation}[1]{%
\setcounter{verticalcompensationrows}{#1}%
\addtocounter{verticalcompensationrows}{-1}%
\vspace*{-\value{verticalcompensationrows}\verticalcompensationlength}%
}

\newcommand{\multirowbt}[3]{%
\multirow{#1}{#2}{\verticalcompensation{#1}#3}%
}

\usepackage{listings}

\newcommand{\goal}[1]{\textcolor{blue}{<***GOAL: #1>}}
\renewcommand{\goal}[1]{}

\definecolor{bluekeywords}{rgb}{0.13,0.13,1}
\definecolor{greencomments}{rgb}{0,0.5,0}
\definecolor{redstrings}{rgb}{0.9,0,0}
\definecolor{grey}{rgb}{0.4,0.4,0.4}

\newcommand{\scripttt}[1]{{\footnotesize\texttt{#1}}}

\usepackage[utf8]{inputenc}
\usepackage{cleveref}
\crefname{section}{§}{§§}
\Crefname{section}{§}{§§}

\begin{document}


\title{Active Access: A Mechanism for High-Performance Distributed Data-Centric Computations}

\author{Maciej Besta}
       \affiliation{Department of Computer Science\\
       ETH Zurich}
       \email{maciej.besta@inf.ethz.ch}
\author{Torsten Hoefler}
       \affiliation{Department of Computer Science\\
       ETH Zurich}
       \email{htor@inf.ethz.ch}

\begin{abstract}
Remote memory access (RMA) is an emerging high-performance programming model that uses
RDMA hardware directly. Yet, accessing
remote memories cannot invoke activities at the target which complicates
implementation and limits performance of data-centric algorithms.
We propose Active Access (AA), a mechanism that integrates well-known
active messaging (AM) semantics with RMA to enable high-performance
distributed data-centric computations. AA supports a new programming
model where the user specifies handlers that are triggered when incoming
{puts} and {gets} reference designated addresses. 
AA is based on a set of extensions to the Input/Output Memory Management
Unit (IOMMU), a unit that provides high-performance hardware support
for remapping I/O accesses to memory.  
%
We illustrate that AA outperforms existing AM and RMA designs,
accelerates various codes such as distributed hashtables or logging
schemes, and enables new protocols such as incremental checkpointing for
RMA.
We also discuss how extended IOMMUs can support a
virtualized global address space in a distributed system that
offers features known from on-node memory virtualization.
We expect that AA can enhance the design of HPC
operating and runtime systems in large computing centers.

\end{abstract}


\maketitle
\pagestyle{plain}

{\vspace{-0.5em}\noindent \textbf{This is a full version of a paper\\ published at ACM ICS'15 under the same title}}

{\vspace{1em}\small\noindent\textbf{Project website:}\\\url{https://spcl.inf.ethz.ch/Research/Parallel\_Programming/Active_Access}}



\section{Introduction}
\label{sec:introduction}

\goal{State that improving performance of current systems is non-trivial}


\goal{Introduce \& advertise RMA \& RDMA, imply they're common and cool}

Scaling on-chip parallelism alone cannot satisfy growing computational demands
of datacenters and HPC centers with tens of thousands of
nodes~\cite{Esmaeilzadeh:2011:DSE:2000064.2000108}.  Remote direct memory
access (RDMA)~\cite{recio2007remote}, a technology that completely removes the
CPU and the OS from the messaging path, enhances performance in such systems.
RDMA networking hardware gave rise to a new class of Remote Memory Access (RMA)
programming models that offer a Partitioned Global Address Space (PGAS)
abstraction to the programmer. Languages such as Unified Parallel C
(UPC)~\cite{upc} or Fortran 2008~\cite{fortran2008}, and libraries such as
MPI-3~\cite{mpi3} or SHMEM implement the RMA principles and enable direct
\emph{one-sided} low-overhead \emph{put} and \emph{get} access to the memories
of remote nodes, outperforming designs based on the Message Passing (MP) model and the associated
routines~\cite{fompi-paper}. 


\goal{Introduce and advertise AM, state an existing research problem}

\sloppy
\emph{Active Messages} (AMs)~\cite{von1992active} are another scheme for
improving performance in distributed environments.  An active message invokes a
handler at the receiver's side and thus AMs can be viewed as lightweight remote
procedure calls (RPC). AMs are widely used in a number of different areas
(example libraries include IBM's DCMF, IBM's PAMI, Myrinet Express (MX), GASNet~\cite{bonachea2002gasnet}, and
AM++~\cite{willcock-amplusplus}). Unfortunately, AMs are limited to message passing and cannot be directly used in RMA
programming.

\goal{Introduce, briefly describe, and motivate AA}

In this work we propose \emph{Active Access} (AA), a mechanism that enhances
RMA with AM semantics. The core idea is that a
remote memory access triggers a user-definable
CPU handler at the target. As we explain in~\cref{sec:motivatingExample}, AA eliminates some of the performance problems
specific to RMA and RDMA.

\goal{Explain (briefly) challenges behind designing AA \& lead to IOMMUs}

Intercepting and processing {puts} or {gets} requires control logic to identify
memory accesses, to decide when and how to run a handler, and to buffer
necessary data.  To preserve all RDMA benefits in AA (e.f., OS-bypass,
zero-copy, others), we propose a hardware-based design that extends the
input/output memory management unit (IOMMU), a hardware unit that supports I/O
virtualization.
IOMMUs evolved from simple DMA remapping devices to units offering advanced
hardware virtualization~\cite{ben-yehuda}. Still, AA shows that many potential
benefits of IOMMUs are yet to be explored.  For example, as we will later show
(\cref{sec:interruptingCPU}), moving the notification functionality from the
NIC to the IOMMU enables high performance communication with the CPU for AA.
Moreover, AA based on IOMMUs can generalize the concept
of virtual memory and enable hardware-supported
virtualization of networked memories
with enhanced data-centric paging capabilities.



\goal{Itemize contributions}


In summary, our key contributions are as follows:
\begin{itemize}[leftmargin=1em]
\item We propose \emph{Active Access} (AA), a mechanism that combines active messages
and RMA to improve the performance of RMA-based applications and systems.
\item We illustrate a detailed hardware design of simple extensions to IOMMUs 
to construct AA.
\item We show that AA enables a new data-centric programming model that
facilitates developing RMA applications.
\item We evaluate AA using microbenchmarks and four large-scale use-cases (a distributed
hashtable, an access counter, a logging system, and fault-tolerant parallel
sort). We show that AA outperforms other communication schemes.
\item We discuss how the IOMMU could enable hardware-based virtualization of remote memories.
\end{itemize}

\subsection{Motivation} 
\label{sec:motivatingExample}

\goal{+ Describe performance problems of RMA}

\begin{figure}[h!]
\centering
\begin{minipage}{.19\textwidth}
  \centering
  \includegraphics[width=1.0\linewidth]{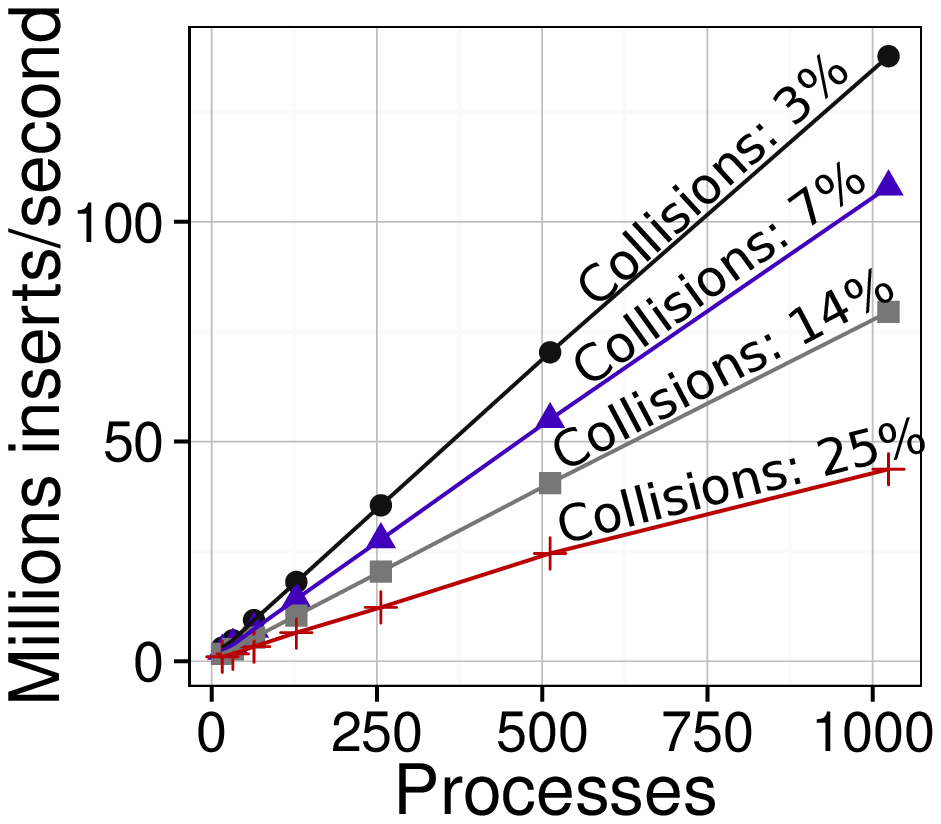}
  \vspace{-2.3em}
\caption{Inserts/s in our RMA hashtable (\cref{sec:motivatingExample})}
  \label{fig:motivation}
\end{minipage}
\hspace*{0.1em}
\begin{minipage}{.24\textwidth}
  \centering
  \includegraphics[width=1\linewidth]{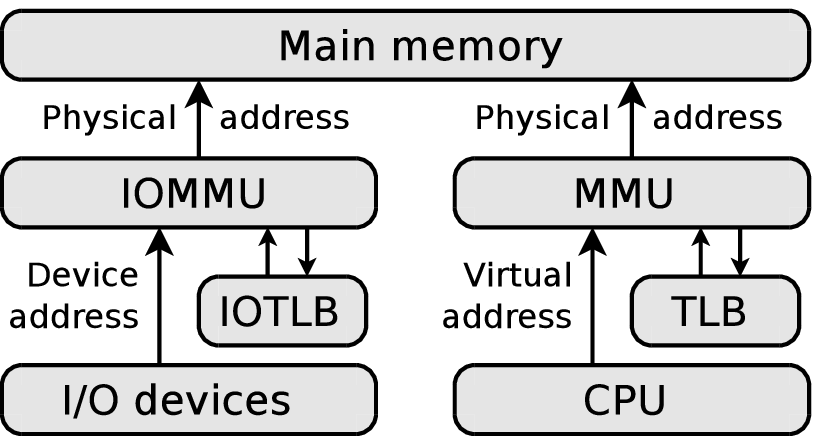}
    \vspace{0.3em}
  \caption{Comparison of the IOMMU and the MMU (\cref{sec:background_iommu})}
  \label{fig:iommu_mmu}
\end{minipage}%
\end{figure}

Consider a distributed hashtable (DHT): RMA programming improves its
performance in comparison to MP 2-10$\times$~\cite{fompi-paper}.
Yet, hash collisions impact performance as handling them requires
to issue many expensive remote atomics (see~\cref{sec:AM_based_applications}).
Figure~\ref{fig:motivation} shows how the performance
varies by a factor
of $\approx$10 with different collision rates.

\goal{+ Motivate the use of IOMMUs}


We will show later (\cref{sec:AM_based_applications}) how AA reduces the 
number of remote accesses from six to \emph{one}. 
Intuitively, the design of AA, based on IOMMU remapping logic,
intercepts memory requests and passes them for direct processing
to the local CPU. Thus, AA combines the benefits of AMs and OS-bypass in RMA
communications.

\vspace{-0.1em}
\section{Background}
\label{sec:background}

\goal{Introduce the Section}

We now briefly outline RMA programming. Then, we discuss the parts of
the IOMMU design (DMA remapping, IOMMU paging) that we later use to design Active Access.

\vspace{-0.1em}
\subsection{RMA Programming Models} 
\label{sec:background_rma}

\goal{+ Describe RDMA \& RMA and show they're popular}

\sloppy
RMA is a programming model in which processes communicate by
directly accessing one another's memories. RMA is typically built on OS-bypass
RDMA hardware to achieve highest
performance. Thus, RMA \emph{put} (writes to
remote memories) and \emph{get} (reads from remote memories) have very low
latencies, and significantly improve performance over
MP~\cite{fompi-paper}. RDMA is
available in virtually all modern networks (e.g., IBM's Cell
on-chip network, InfiniBand~\cite{IBAspec}, IBM PERCS, iWARP, and RoCE).  In addition, numerous
existing languages and libraries based on RMA such as UPC, Titanium,
Fortran 2008, X10, Chapel, or MPI-3.0 RMA are actively developed
and offer unique features for parallel programming. Consequently, the
number of applications in the RMA model is growing rapidly.

\goal{+ Introduce the naming for sources and targets}

Here, we use \emph{source} or \emph{target} to
refer to a process that issues or is targeted by
an RMA access. We always use \emph{sender}
and \emph{receiver} to refer to processes that exchange messages.

\vspace{-0.2em}
\subsection{IOMMUs}
\label{sec:background_iommu}

\goal{+ Introduce and briefly describe IOMMUs}

IOMMUs are located between peripheral devices and main memory and
can thus intercept any I/O traffic. Like the well-known memory
management units (MMUs), they can be programmed to translate device
addresses to physical host addresses. Figure~\ref{fig:iommu_mmu}
compares MMUs and IOMMUs. An IOMMU can virtualize the view of
I/O devices and control access rights to memory pages. 

\begin{figure*}[ht]
\centering
\includegraphics[width=1.0\textwidth]{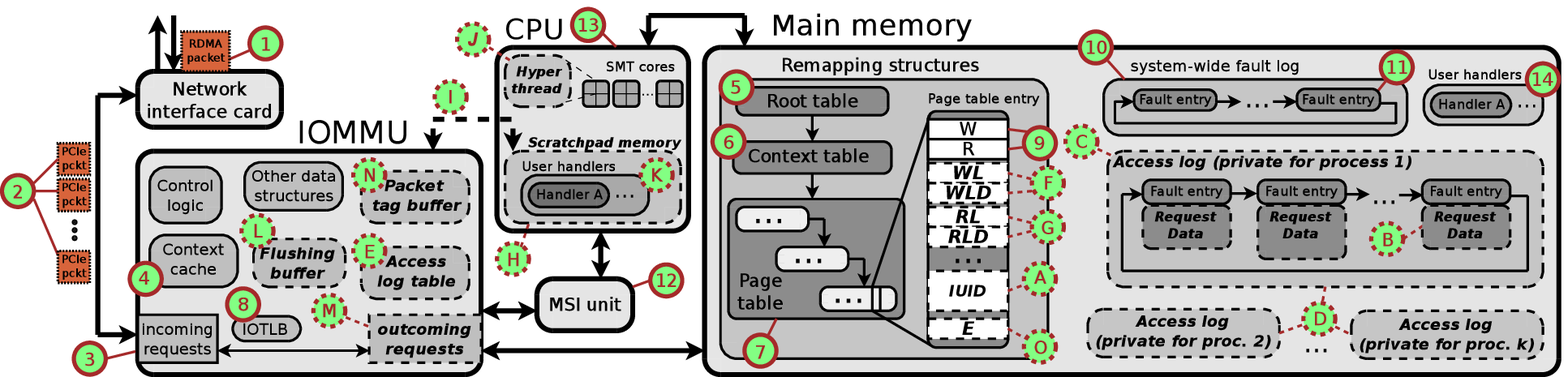}
\caption{The overview of the IOMMU and the cooperating devices. The proposed
extensions are marked with dashed edges and bold-italic text. Solid
circles with numbers (\protect\includegraphics[scale=0.65,trim=0 3.5 0
0]{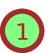} - \protect\includegraphics[scale=0.65,trim=0 3.5 0
0]{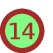}) indicate the specific steps discussed in detail in~\cref{sec:examplePageFault}. Dashed circles
(\protect\includegraphics[scale=0.65,trim=0 3.5 0 0]{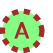} -
\protect\includegraphics[scale=0.65,trim=0 3.5 0 0]{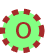}) are extensions
pointed out in~\cref{sec:examplePageFault}-\cref{sec:implIssues}.}

\label{fig:iommu_fault_action}
\end{figure*}

\goal{++ Show that IOMMUs are common and conclude we want them}

All major hardware vendors such as IBM, Intel, AMD, Sun, and ARM offer
IOMMU implementations to support virtualized environments; 
Table~\ref{tab:iommus_compar} provides an overview. We conclude that
IOMMUs are a standard part of modern computer architecture and the
recent growth in virtualization for cloud computing ensures that
they will remain important in the future. 
However, IOMMUs are a relatively new concept with many unexplored
opportunities. Their ability to intercept any memory access and
provide full address space virtualization can be the basis for
many novel mechanisms for managing global (RDMA) address spaces. 

\goal{++ Say we use VT-d, still show our design is generic}

To be as specific as possible, we selected Intel's IOMMU technology~\cite{intel_vtd} to
explain concepts of generic IOMMUs. Other implementations vary in
some details but share the core features (per-page protection, DMA remapping,
etc.).

\goal{++ Describe the remapping of memory accesses}


\textsf{\textbf{DMA Remapping }} 
DMA remapping is the IOMMU function that we use extensively to design AA.
The IOMMU remapping logic allows any I/O device to be assigned to its
own private subset of host physical memory that is isolated from accesses
by other devices. To achieve this, IOMMUs utilize three types of
remapping structures (all located in main memory): \emph{root-entry}
tables, \emph{context-entry} tables, and IOMMU \emph{page tables}. The first two
are used to map I/O devices to device-specific page tables.
To improve the access time, the remapping hardware maintains several 
caches such as the \emph{context-cache} (device-to-page-table
mappings) and the \emph{I/O Translation Lookaside Buffer} (IOTLB) 
(translations from device addresses to host physical addresses).

\begin{table}
\setlength{\tabcolsep}{1.7pt}
\footnotesize
\centering
\begin{tabular}{>{\centering\arraybackslash}m{1.2cm} p{7cm}}
\toprule
\textbf{\textsf{Vendor}}&\textbf{\textsf{IOMMU and its application}}\\
\midrule
\multirowbt{3}{*}{\textsf{AMD}}&\textsf{GART~\cite{amd_gart}: address translation for the use by AGP}\\
&\textsf{DEV~\cite{ben-yehuda}: memory protection}\\
&\textsf{AMD IOMMU~\cite{amd_iommu}: address translation \& memory protection}\\
\midrule
\multirowbt{5}{*}{\textsf{IBM}}&\textsf{Calgary PCI-X bridge~\cite{ben-yehuda}: address translation, isolation}\\
&\textsf{DART~\cite{ben-yehuda}: address translation, validity tracking}\\
&\textsf{IOMMU in Cell processors~\cite{ben-yehuda}: address translation, isolation}\\
&\textsf{IOMMU in POWER5~\cite{Armstrong:2005:AVC:1148882.1148885}: hardware enhanced I/O virtualization}\\
&\textsf{TCE~\cite{ibm_tce}: enhancing I/O virtualization in pSeries 690 servers}\\
\midrule
\textsf{Intel}&\textsf{VT-d~\cite{intel_vtd}: memory protection, address translation}\\
\midrule
\textsf{ARM}&\makecell[l]{\textsf{CoreLink SMMU~\cite{arm_iommu}: memory management in System-on-Chip}\\\textsf{ (SoC) bus masters, memory protection, address translation}}\\
\midrule
\textsf{PCI-SIG}&\textsf{IOV \& ATS~\cite{pci_iommu}: address translation, memory protection}\\
\midrule
\textsf{Sun}&\textsf{IOMMU in SPARC~\cite{sparc_iommu}: address translation, memory protection}\\
\midrule
\textsf{SolarFlare}&\textsf{IOMMU in SF NICs~\cite{openonload}: address translation, memory protection}\\
\bottomrule
\end{tabular}
\caption{An overview of existing IOMMUs (\cref{sec:background_iommu}).}
\label{tab:iommus_compar}
\vspace{-2.8em}
\end{table}


\goal{+++ Describe IOMMU page tables}

\textsf{\textbf{Page Tables \& Page Faults }}
IOMMU page tables allow to manage host physical memory hierarchically; they are similar to standard MMU page tables
(still, MMU and IOMMU
page tables are independent).
A 4-level table
allows 4KB page granularity on 64 bit machines (superpages of various sizes are
also supported).

\goal{+++ Describe IOMMU page fault mechanism}

IOMMU page tables implement a page fault mechanism similar to MMUs.
Every page table entry (PTE) contains two protection bits, \texttt{W} and
\texttt{R}, which indicate whether the page is writable and readable, respectively.
Any access that violates the
protection conditions is blocked by the hardware and a \emph{page fault}
is generated, logged, and reported to the OS. The IOMMU logs the
fault information using special registers and in-memory \emph{fault
logs}. The OS is notified using Message Signal Interrupts (MSI).  Every
{page fault} is logged as a fixed-sized \emph{fault entry} that
contains the fault metadata (the
address of the targeted page, etc.); the data being transferred is discarded.
We will later extend this mechanism to log active accesses and their data and to bypass the OS.

\section{The Active Access Mechanism}
\label{sec:aa_prot}

\goal{Briefly describe AA using an analogy to AM}

Active Access combines
the benefits of RMA and AMs. 
AMs enhance the message passing model by allowing messages to actively integrate into
the computation on the receiver side. In RMA, processes communicate by
accessing remote memories instead of sending messages. Thus, an
analogous scheme for RMA has to provide the \emph{active} semantics for
both types of remote operations; {puts} and {gets} become
\emph{active puts} (AP) and \emph{active gets} (AG), respectively.
%
%


Listing~\ref{lst:semantics} shows the interface of AM and AA.
An active message sent to a process \texttt{receiver\_id} carries \texttt{arguments}
and \texttt{payload} that will be used by a handler identified by a
pointer \texttt{hlr\_addr}.
In AA, the user issues puts and gets at \texttt{trgt\_addr} in the address space of a process \texttt{trgt\_id}. No handler address is
specified. Instead, we enable the user to associate an arbitrary page of data with a selected handler and with a set of additional actions (discussed in~\cref{sec:exampleExtensionsPuts} and~\cref{sec:exampleExtensionsGets}). When a put or a get touches such a page, it becomes \emph{active}: first, it may or may not finalize its default memory effects (depending on the specified actions); second, \emph{both} its metadata \emph{and} data are ultimately processed by the associated handler. AA is fully transparent to RMA and, as Listing~\ref{lst:semantics} shows, it entails no changes to the traditional interface.

\begin{lstlisting}[float=h,xleftmargin=1em,caption=Interface of Active Messages and Active Access
(\cref{sec:aa_prot}),label=lst:semantics, mathescape=true]
/***************** interface of AM ****************/
void send_active_message(ptr hlr_addr, void* arguments, void* payload, int receiver_id) { ... }
/***************** interface of AA ****************/
void put(void* trgt_addr, void* data, int trgt_id) { 
	/* Attempt to copy $data$ from the local memory into
	the memory location $trgt\_addr$ of a process $trgt\_id$.*/
}
void get(void* trgt_addr, void* l_addr, int trgt_id) {
	/* Attempt to fetch the data from the memory location
	$trgt\_addr$ of a process $trgt\_id$ to $l\_addr. $*/
}
void assoc_page(void* addr, void* act, int hlr_id) {
	/* Associate a page at $addr$ with actions $act$ and
	with a handler identified by the id $hlr\_id$. */
\end{lstlisting}


We now show how to extend IOMMUs to implement the above AA interface
and to enable active puts and gets. From
now on, we will focus on designs based on PCI Express (PCIe)~\cite{pci3_spec}.
We first describe the interactions between an RDMA request and current
IOMMUs. The numbers in circles
(\includegraphics[scale=0.8,trim=0 3 0 0]{1.eps}~-~\includegraphics[scale=0.8,trim=0 3 0 0]{14.eps}) refer to the
corresponding numbers in Figure~\ref{fig:iommu_fault_action}.

\begin{figure*}
\centering
 \subfloat[An {active put} (\cref{sec:exampleExtensionsPuts})]{
  \includegraphics[width=0.48\textwidth]{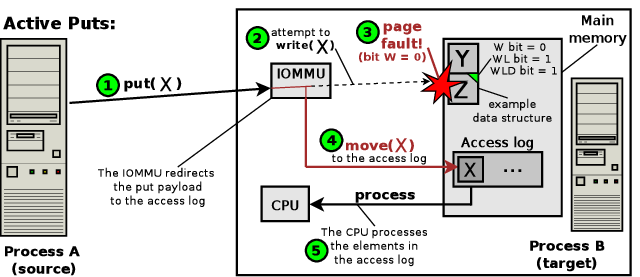}
  \label{fig:remoteCOW}
 }\hfill
 \subfloat[An {active get} (\cref{sec:exampleExtensionsGets})]{
  \includegraphics[width=0.48\textwidth]{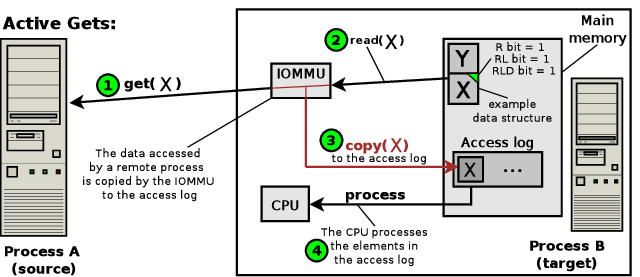}
  \label{fig:remoteCOR}
 }
 \caption{{Active puts and gets}. Here, the numbers in circles are independent of the numbering in Figure~\ref{fig:iommu_fault_action}.}
\label{fig:mb_bb}
\end{figure*}

\subsection{State-of-the-art IOMMU Processing Path}
\label{sec:examplePageFault}

%

\goal{+ Describe the remapping mechanism till the page fault}

Consider an RDMA {put} or a {get} that is issued by a remote process.
First, the local NIC receives the RDMA packet~\includegraphics[scale=0.8,trim=0 3 0 0]{1.eps}. The NIC attempts to access
the main memory with DMA and thus it generates appropriate PCIe packets (one
or more depending on the type of the PCIe transaction~\cite{pci3_spec})~\includegraphics[scale=0.8,trim=0 3 0 0]{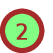}. Each packet is intercepted
by the IOMMU~\includegraphics[scale=0.8,trim=0 3 0 0]{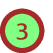}. First, the
IOMMU resolves the mapping from the device to its page table (using the
packet header~\cite{intel_vtd}). Here, the IOMMU uses
the context cache~\includegraphics[scale=0.8,trim=0 3 0 0]{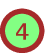} or, in
case of a cache miss, it walks the remapping tables~\includegraphics[scale=0.8,trim=0 3 0 0]{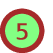}-\includegraphics[scale=0.8,trim=0 3 0 0]{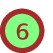}. Finally, the IOMMU obtains
the location of the specific page-table~\includegraphics[scale=0.8,trim=0 3 0
0]{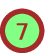}. The IOMMU then resolves the mapping from a device
address to a physical address using the IOTLB~\includegraphics[scale=0.8,trim=0 3 0
0]{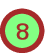} or, in the event of a miss, the page-table~\includegraphics[scale=0.8,trim=0 3 0 0]{7.eps}. When it finds the target
PTE, it checks its protection bits \texttt{W} and \texttt{R}~\includegraphics[scale=0.8,trim=0 3 0
0]{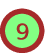}. The next steps depend on the request type. For {puts}, if
\texttt{W=1}, the value is simply written to
the target location. If \texttt{W=0}, the IOMMU raises a page fault and does not
modify the page. For {gets}, if \texttt{R=1},
the request returns the accessed value to the NIC. If
\texttt{R=0}, the IOMMU raises a page fault and does not return the value.

\goal{+ Describe the remainder of the remapping mechanism}

Upon a page fault, the IOMMU tries to record the fault information
({fault entry}) in the system-wide {fault log}~\includegraphics[scale=0.8,trim=0
3 0 0]{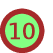} (implemented as an in-memory ring buffer). In case
of an overflow (e.g., if the OS does not process the recorded entries fast
enough) the {fault entry} is not recorded. If the {fault
entry} is logged~\includegraphics[scale=0.8,trim=0 3 0 0]{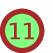},
the IOMMU interrupts the CPU~\includegraphics[scale=0.8,trim=0 3 0 0]{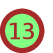} with MSI~\includegraphics[scale=0.8,trim=0 3 0
0]{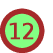} to run one of the
specified handlers~\includegraphics[scale=0.8,trim=0 3 0 0]{14.eps}	.

\goal{+ Guide to active puts}


We now analyze the extensions that enable {active puts/gets}
(symbols~\includegraphics[scale=0.8,trim=0 3 0 0]{A.eps}~-~\includegraphics[scale=0.8,trim=0 3 0 0]{O.eps} refer to the related
symbols in Figure~\ref{fig:iommu_fault_action}). Our goal is to enable the
IOMMU to multiplex intercepted {accesses} among processes, buffer them in designated memory locations,
and pass them for processing to a CPU.
%

\subsection{Processing the Intercepted Data}
\label{sec:multiplexing-the-data}

 

In the original IOMMU design the fault log~\includegraphics[scale=0.8,trim=0 3 0 0]{10.eps} is shared by all the processes
running on the node where the IOMMU resides; a potential performance bottleneck. In addition, the IOMMU does not enable multiplexing
the data coming from the NIC across the processes and handlers, limiting
performance in multi/manycore environments.
Finally, the data of a blocked RDMA put is
lost as the fault log entry only records the address (see steps~\includegraphics[scale=0.8,trim=0 3 0
0]{9.eps}-\includegraphics[scale=0.8,trim=0 3 0 0]{10.eps}).
%
%
To alleviate these issues, we propose to enhance the design of the IOMMU and its page tables to enable a data-centric
multiplexing mechanism in which \emph{the PTEs themselves guide the incoming requests to be recorded
in the specified logging data structures and processed by the designated user-space handlers}.


We first add a programmable field
\emph{IOMMU User Domain ID} (IUID) to every IOMMU PTE~\includegraphics[scale=0.8,trim=0 3 0 0]{A.eps}.
This field enables \emph{associating} pages with user domains. The OS and the NIC can ensure that one IUID is associated with at most one local process,
similarly to Protection Domains in RDMA~\cite{recio2007remote}. To add IUID we use bits 52-61
of IOMMU PTEs (ignored by the current IOMMU hardware~\cite{intel_vtd}).
We can store
$2^{10}$ domains on each node; enough to fully utilize, e.g., BlueGene/Q (64 hardware threads/node) or Intel Xeon Phi (256 hardware threads/chip).
Second, our extended IOMMU logs {\emph{both}} the 
{generated fault entry {\emph{and}} the carried data~\includegraphics[scale=0.8,trim=0 3 0
0]{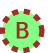}} to the \emph{access log}, a new in-memory circular ring buffer~\includegraphics[scale=0.8,trim=0 3 0
0]{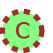}. A process can have multiple private IUIDs/access logs located in its
address space~\includegraphics[scale=0.8,trim=0 3 0 0]{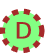}.
Third, the IOMMU maintains the \emph{access
log table}~\includegraphics[scale=0.8,trim=0 3 0 0]{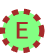}, a simple
internal associative data structure with tuples \texttt{(IUID,base,head,tail,size)}.
One entry maps an IUID to three physical addresses (the base, the head, and the tail pointer) and the size of the respective access log ring buffer.
The access log table is implemented as content addressable
memory (CAM) for rapid access and it can be programmed in the same way as other Intel IOMMU
structures~\cite{intel_vtd}.
%

\subsection{Controlling Active Puts (APs)}
\label{sec:exampleExtensionsPuts}




\goal{++ Describe the active bits that control the logging}

Active puts enable redirecting data coming from the NIC
and/or related metadata to a specified access log.
Figure~\ref{fig:remoteCOW} illustrates active puts in more detail.
Two additional PTE bits control the logging of fault
entry and data: \texttt{WL} (Write Log)
and \texttt{WLD} (Write Log Data)~\includegraphics[scale=0.8,trim=0 3 0 0]{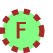}.
If \texttt{WL=1} then the IOMMU logs the fault
entry for the written page and if \texttt{WLD=1} then the IOMMU logs both the fault entry
and the data. 
%
The flags \texttt{W}, \texttt{WL} and \texttt{WLD} 
are independent. For example, an active put page
is marked as \texttt{W=0},
\texttt{WL=1}, \texttt{WLD=1}.
The standard way, in which IOMMUs manage
faults triggered by writes, is defined by the
values \texttt{W=0}, \texttt{WL=1}, \texttt{WLD=0}.

\subsection{Controlling Active Gets (AGs)}
\label{sec:exampleExtensionsGets}

\goal{++ Describe Active Gets}

Active gets enable the IOMMU to log a copy of the
remotely accessed data locally.
When a {get} succeeds and the returned data is flowing from the main memory to the NIC, it
is replicated by the IOMMU and saved in the {access log} (see
Figure~\ref{fig:remoteCOR}).
%
Similar to active puts, two additional PTE bits control the logging behavior of
such accesses: \texttt{RL} (Read Log) and \texttt{RLD} (Read Log Data)
\includegraphics[scale=0.8,trim=0 3 0 0]{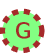}.  If \texttt{RL=1}
then the IOMMU logs the fault entry for the read page. If
\texttt{RLD=1}, the IOMMU logs the fault entry \emph{and} the returned data. 



\goal{++ Say how we can implement these bits}

The proposed control bit extensions for active puts and active gets can easily be implemented in practice. For
example, bits 7-10 in the Intel IOMMU PTEs are ignored~\cite{intel_vtd}. These bits can be used to store
\texttt{WL}, \texttt{WLD}, \texttt{RL}, and \texttt{RLD}.

\subsection{Interactions with the Local CPU}
\label{sec:interruptingCPU}

\goal{+ Motivate and describe the extensions for CPU communication}

%

Finally, the IOMMU has to notify the CPU to run a handler to process the logs.
Here, we discuss interrupts/polling
and we propose a new scheme where
the IOMMU directly accesses the CPU, bypassing the main memory.


\textsf{\textbf{Interrupts }}
%
Here, one could use a high-performance MSI wakeup mechanism analogous to the
scheme in InfiniBand~\cite{IBAspec}. The developer
specifies conditions for triggering interrupts (when the amount of
free space in an access log is below a certain threshold, or at pre-determined intervals).
If the access log is sufficiently large and messages are pipelined then the
interrupt latency may not influence the overall
performance significantly (cf.~\cref{sec:evaluation}).
%



\textsf{\textbf{Polling }}
As the IOMMU inserts data directly into a user
address space, processes can monitor the access log head/tail pointers
and begin processing the data when required.
Polling can be done either directly by the user, or by a runtime
system that runs the handlers
transparently to the user.
%

%


\textsf{\textbf{Direct CPU Access }}
This mechanism is motivated by the architectural trends to place
scratchpad memories on processing units, a common practice in
today's NVIDIA GPUs~\cite{patterson2009top} and 
several multicore architectures~\cite{kim2014wcet}.
One could add a scratchpad to the CPU~\includegraphics[scale=0.8,trim=0 3 0 0]{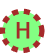}, connect it
directly with the IOMMU~\includegraphics[scale=0.8,trim=0 3 0 0]{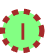}, and place
the head/tail pointers in it.
A dedicated hyperthread~\includegraphics[scale=0.8,trim=0 3 0 0]{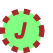} polls the pointers and runs the handlers if a free entry is available~\includegraphics[scale=0.8,trim=0 3 0 0]{14.eps}.
If the size of the handler code is small, it can also be placed in the scratchpad, further reducing the number of memory accesses~\includegraphics[scale=0.8,trim=0 3 0 0]{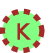}.


The IOMMU and the CPU also have to synchronize while processing
the access log. This can be done with a simple lock for mutual access.
The IOMMU and the CPU could also synchronize with
the pointers from the access log table.

%

\subsection{Consistency Model}
\label{sec:consistency}


\goal{Introduce and provide the intuition behind active flushes}

We now enhance AA to enable a weak consistency model similar to MPI-3
RMA~\cite{mpi3-rma-overview}.
In RMA, a blocking flush synchronizes nonblocking puts/gets.
%
%
%
In AA, we use an \emph{active flush} 
(\texttt{flush(int target\_id)})
to enforce the completion of active accesses issued by the calling process
and targeted at \texttt{target\_id}. 
%
%
One way to implement active flushes could be to issue an active \texttt{get} targeted
at a special designated \emph{flushing page} in the address space of \texttt{target\_id}.
The IOMMU, upon intercepting this get, would wait until the CPU processes the related access log
and then it would finish the \texttt{get} to notify the source that the accesses are committed.


\textsf{\textbf{Extending the IOMMU }}
We add an IOMMU internal data structure called the \emph{flushing buffer}~\includegraphics[scale=0.8,trim=0 3 0 0]{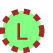} to store tuples
\texttt{(address,IUID,active,requester-ID,tag)}, where \texttt{address} is the address of the flushing page,
\texttt{active} is a binary value initially set to \texttt{false}, and \texttt{requester-ID,tag}
are values of two PCIe packet fields with identical names; they are initially zeroed
and we discuss them later in this section.
The flushing buffer is implemented as CAM for rapid access.


\textsf{\textbf{Selecting a Flushing Page }}
%
%
%
To enable a selection of a flushing page, the system could reserve a high virtual
address for this purpose.
We then add a respective entry (with the selected address and
the related IUID) to the flushing buffer.
%
%
%

%
%



\textsf{\textbf{Finishing an Active Flush }}
%
%
%
%
%
The IOMMU intercepts the issued get,
finds the matching entry in the flushing buffer, 
sets \texttt{active=true},
copies the values of the \texttt{tag} and \texttt{requester-ID} PCIe fields
to the matching entry, and discards the get.
%
%
Processing of a targeted access log is then initiated with any scheme from~\cref{sec:interruptingCPU},
depending on user's choice.



\textsf{\textbf{Alternative Mechanism }}
The proposed consistency mechanism
sacrifices one page from the user virtual address space.
To alleviate this, we offer a second scheme similar to
the semantics offered by, e.g., GASNet~\cite{bonachea2002gasnet}.
%
%
Here, AA does not guarantee any consistency. Instead,
it allows the user to develop the necessary consistency by issuing a \emph{reply} (implemented as an active put) from within the handler.
This reply informs which elements
from the access log have been processed.
%
%
To save bandwidth, replies can be batched.
The reply would be targeted at a designated page (with bits \texttt{W=0},
\texttt{WL}=1, \texttt{WLD}=1) with an \texttt{IUID}
pointing to a designated access log.
The user would poll the log
and use the replies to enforce an arbitrary desired consistency.




\textsf{\textbf{Mixing AA/RMA Accesses }}
At times, mixing AA and RMA puts/gets
may be desirable (see~\cref{sec:distHT}).
The consistency of such a mixed scheme can be managed
with active and traditional RMA flushes: these two
calls are orthogonal.
%
Completing pending AA/RMA
accesses is enforced with AA/RMA flushes, respectively.

\subsection{Hardware Implementation Issues}
\label{sec:implIssues}

\goal{+ Say why we care about control flow and ordering issues}

We now describe solutions to several PCIe and RDMA control flow, ordering, and
backward compatibility issues in the proposed extensions. If the reader is not
interested in these details then they may skip this part and proceed
directly to Section~\ref{sec:AM_based_applications}.
Numbers and capitals in circles refer to Figure~\ref{fig:iommu_fault_action}.
%

\textsf{\textbf{Logging Data of PCIe Write Requests }}
\label{sec:putPCIePackets}
\goal{++ Explain how we manage flow of PCIe packets}
Every RDMA {put} is translated into one or more PCIe write
requests flowing from the NIC to the main memory. The ordering rules for
\emph{Posted Requests} from the PCIe specification~\cite{pci3_spec} (§ 2.4.1,
entry~A2a) ensure that the packets for the
same request arrive in order. Consequently, such packets can simply be appended to the log.

\textsf{\textbf{Logging Data of PCIe Read Requests }}
\label{sec:readPCIePackets}
\goal{++ Motivate and describe more complex extensions for AG}
A PCIe read transaction consists of one read request (issued by the NIC)
and one or more read replies (issued by the memory controller). The
IOMMU has to properly match the incoming and outgoing PCIe packets. For this, we
first enable the IOMMU to intercept PCIe packets flowing back to
the NIC \includegraphics[scale=0.8,trim=0 3 0 0]{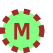} (standard
IOMMUs process only incoming memory accesses). Second, we
add the \emph{packet tag buffer} \includegraphics[scale=0.8,trim=0 3
0 0]{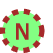} (implemented as CAM) to the IOMMU to
temporarily maintain information about PCIe packets.
%
%
%
The IOMMU would add the transaction tags of incoming PCIe read
requests that access a page where \texttt{RLD=1} to the tag buffer. PCIe
read reply packets are then matched against the buffer and logged if
needed.
%
%
%
%
\goal{+++ Say why we need to use this complex tag buffer}
%
%
We require the tag buffer as PCIe read replies only contain
seven lower bits of the address of the accessed memory region (see § 2.2.9 in~\cite{pci3_spec}), preventing the IOMMU from matching incoming requests with replies. 
The PCIe standard also ensures ordering of read replies
(see § 2.3.1.1 in~\cite{pci3_spec}).

%
%
%

\textsf{\textbf{Order of PCIe Packets from Multiple Devices }}
\goal{++ Describe how we maintain ordering of packets for interleaved accesses}
%
%
The final ordering issue concerns multiple RMA {puts} or {gets}
concurrently targeted at the same IOMMU. If several multi-packet accesses originate
from different devices then the IOMMU may observe an incoming arbitrary
interleaving of PCIe packets. To correctly reassemble the packets in the access log, we
extend the tag buffer so that it also stores pointers into the access
log. Upon intercepting the first PCIe packet of a new PCIe transaction,
the IOMMU inserts a tuple \texttt{(transaction-tag, tail)}
into the tag buffer. Then, the IOMMU records the packet in the access log, and adds the
size of the \emph{whole} PCIe transaction to the tail pointer. Thus, if
some transactions interleave, the IOMMU leaves ``holes'' in the {access
log} and fill these holes when appropriate PCIe packets
arrive\footnote{\footnotesize PCI Express 3.0 Specification limits the PCIe
transaction size to 4KB. Thus, the maximum size of an active put or get
also amounts to 4 KB. This limitation can be easily overcome in future
PCIe systems.}. The IOMMU removes an entry from the buffer after
processing the last transaction packet. To ensure that the CPU
only processes packets with no holes, the IOMMU increments \texttt{tail}
or sets appropriate synchronization variables
only when each PCIe packet of the next transaction is recorded in the respective
access log.

\goal{++ Describe how we handle potential overflows}

\textsf{\textbf{Control Flow }}
IOMMUs, unlike MMUs, cannot suspend a remote process and thus buffers
may overflow if they are not emptied fast enough. To avoid data loss, we
utilize the backpressure mechanism of the PCIe transaction layer
protocol (TLP) as described in § 2.6.1.\ in~\cite{pci3_spec}. This
will eventually propagate through a reliable network and block the
sending process(es). Issues such as head of line blocking and deadlocks
are similar to existing reliable network technologies and require
efficient programming at the application layer (regular emptying of the
queues). Head of line blocking can also be avoided by
dropping packets and retransmission~\cite{portals4}.




%

\textsf{\textbf{Support for Legacy Codes }}
Some codes may rely on the default IOMMU behavior to buffer the metadata in the default fault log~\includegraphics[scale=0.8,trim=0 3 0 0]{10.eps}.
To cover such cases, we add the \texttt{E} bit~\includegraphics[scale=0.8,trim=0 3 0
0]{O.eps} to IOMMU PTEs to determine if the page fault is recorded in the
fault log (\texttt{E=0}) or in one of the access
logs (\texttt{E=1}).

\section{Active Access Programming}
\label{sec:AM_based_applications}

%
%

\goal{Say what AA is and how it enhances performance}

We now discuss
example RMA-based codes that leverage AA.
%
AA improves the application performance
by reducing the amount of communication and remote synchronization, enhancing 
locality~\cite{tate2014programming}.
First, it reduces the number of puts, gets, and remote atomic operations in
distributed data structures and other codes that perform complex
remote memory accesses. For example,
enqueueing an element into a remote queue costs at least two remote
accesses (atomically get and increment the tail pointer and put the
element). With AA, this would be a simple put to the list address and a
handler that inserts the element; our DHT example is very similar. 
%
%
Second, as handlers are executed by local cores, the
usual on-node synchronization schemes are used with no need
to issue expensive remote synchronization calls, e.g., remote locks.

\begin{figure*}
\centering
\noindent\begin{minipage}{.48\textwidth}
\begin{lstlisting}[caption=Insert in the traditional RMA-based DHT,label=lst:insert_rma, mathescape=true]
/* Volume is a structure that contains the fields:
$owner$: the id of the volume owner; $vol\_size$: volume size,
$elems[]$: the table + the overflow heap; each cell contains two subfields: $elem$ (the actual value) and $ptr$ (the pointer to the next element),
$next\_free\_cell$: a ptr to the next free cell in the heap,
$last\_ptr[]$: pointers to the most recent elements */

void insert(int elem, Volume v) {//put $elem$ into volume $v$
	int pos = hash(elem);	//get the position of $elem$ in $v$
	if(cas(elem,$\emptyset$,v.elems[pos].elem,v.owner) != $\emptyset$) {
		int free_cell = fao(SUM,1,v.next_free_cell,v.owner);
		if(free_cell>=v.vol_size) {/*an overflow - resize*/}
		rma_put(elem,v.elems[free_cell].elem,v.owner);
		rma_flush(v.owner);
		int prev_ptr=fao(REPLACE,free_cell,v.last_ptr[pos],  v.owner); 
		if(cas(free_cell,$\emptyset$,v.elems[pos].ptr,v.owner) != $\emptyset$) {
			rma_put(free_cell,v.elems[prev_ptr].ptr,v.owner);
			rma_flush(v.owner); } } }
\end{lstlisting}
\end{minipage}\hfill
\begin{minipage}{.48\textwidth}
\begin{lstlisting}[caption=Insert in the AA-based DHT,label=lst:insert_aa, mathescape=true]
void insert(int elem, Volume v) {
  put(elem, v.elems[hash(elem)].elem, v.owner); 
}

void insert_handler(Access_log log) {
	while(log.tail != log.head) {
		local_insert(*log.tail); log.tail += sizeof(int);
    	if(log.tail == log.base + log.size) {
        	log.tail = log.base;
} } }

void local_insert(int elem) {//$lv$ is the local DHT volume
	int pos = hash(elem);	//get the position of $elem$ in $lv$
	if(cas(elem,$\emptyset$,lv.elems[pos].elem) != $\emptyset$) {
		int free_cell = fao(SUM,1,lv.next_free_cell);
		if(free_cell>=lv.vol_size) {/*an overflow - resize*/}
		lv.elems[free_cell].elem = elem;
		int prev_ptr=fao(REPLACE,free_cell,lv.last_ptr[pos]); 
		if(cas(free_cell,$\emptyset$,lv.elems[pos].ptr) != $\emptyset$) {
			lv.elems[prev_ptr].ptr = free_cell; } } }
\end{lstlisting}
\end{minipage}
\label{fig:dht-listings}
\end{figure*}

\subsection{Designing Distributed Hash Tables}
\label{sec:distHT}

\goal{+ Describe the hashtable}

DHTs are basic data structures that are used to construct distributed key-value
stores such as Memcached~\cite{fitzpatrick2004distributed}.
%
%
In AA, the DHT is open and each process manages its part called the local
\emph{volume}. The volume consists of a table of elements and an overflow heap
for elements with hash collisions. Both the table and the heap are implemented
  as fixed-size arrays. To avoid costly array traversals, pointers to most
  recently inserted items and to the next free cells are stored along with the
  remaining data in each local volume.

\goal{+ Describe the insert function based on RMA}

Due to space constraints we discuss inserts and then we only briefly outline lookups and deletes.
In RMA, inserts are based on
atomics~\cite{schweizer2015evaluating} (Compare-and-Swap and Fetch-and-Op, denoted as \texttt{cas} and \texttt{fao}), RMA puts (\texttt{rma\_put}) and RMA flushes (\texttt{rma\_flush}); see Listing~\ref{lst:insert_rma}.
For simplicity we assume that atomics are blocking.
The semantics of CAS are as follows: \texttt{int cas(elem, compare, target, owner)};
if \texttt{compare} == \texttt{target} then \texttt{target} is changed to \texttt{elem}
and its previous value is returned.
For FAO we have \texttt{int fao(op, value, target, owner)}; it applies an atomic operation \texttt{op}
to \texttt{target} using a parameter \texttt{value}, and returns \texttt{target}'s previous value.
In both \texttt{cas} and \texttt{fao}, \texttt{owner} is the id of the process that owns
the targeted address space.
The semantics for \texttt{rma\_put} and \texttt{rma\_flush} are the same
as for AA puts and flushes (cf.~\cref{sec:aa_prot},~\cref{sec:consistency}).
$\emptyset$ indicates that the specific array cell is empty.
To insert \texttt{elem} we first issue a
\texttt{cas} (line~9). Upon a collision we
acquire a new element in the overflow heap (line 10). We then insert \texttt{elem} into
the new position (lines~12-13), update the respective
last pointer and the next pointer of the previous element in the heap (lines~14-17).

\goal{++ Describe the insert based on AA}

\textsf{\textbf{Implementation of Inserts with Active Puts }}
We now accelerate inserts with AA.
We present the multi-threaded code in Listing~\ref{lst:insert_aa}.
The inserting process calls \texttt{insert} (lines 1-3).
The PTEs of the hash table data are marked with \texttt{W=0},
\texttt{WL}=1, \texttt{WLD}=1; thus,
the metadata and the data from the put
is placed in the access log.
The CPU then (after being interrupted or by polling
the memory/scratchpad) executes \texttt{insert\_handler} to insert the elements 
into the local volume (lines~5-10). Here, we assume that a thread owns one access
log and that the size of the access log is divisible by \texttt{sizeof(int)}.
Elements are inserted with \texttt{local\_insert}, a function similar to \texttt{insert}
from Listing~\ref{lst:insert_rma}. The difference is that each call is local (consequrntly, we skip the
\texttt{lv.owner} argument).


\goal{++ Describe the synchronization of inserts}

\textsf{\textbf{Synchronization }}
AA handlers are executed by the local CPU, thus, \texttt{local\_insert}
requires no synchronization with remote processes. 
In our code we use local atomics, however, other simple {local}
synchronization mechanisms (e.g., locks or hardware transactional memory)
may also be utilized.

\textsf{\textbf{Consistency }}
The proposed DHT is loosely consistent.
For implementing any other consistency (e.g., sequential consistency)
one can use either active flushes or enforce the required consistency using replies from within the handler.


\textsf{\textbf{Lookups }}
Contrary to inserts, lookups do not generate hash collisions that entail
multiple memory accesses.
Thus, we propose to implement a lookup as a single \emph{traditional} RMA get,
similarly to the design in FaRM by Dragojevic et al.~\cite{dragojevic2014farm}.
For this, we mark the PTEs associated with the hashtable data as
\texttt{R=1}, \texttt{RL=0}, \texttt{RLD=0}. Here, we assume that 
DMA is cache coherent (true on, e.g., Intel x86~\cite{dragojevic2014farm})
and that RMA gets are aligned. As the DHT elements are word-size integers,
a get is atomic with respect to any concurrent accesses from Listing~\ref{lst:insert_aa}.
Consistency with other lookups and with inserts can be achieved with RMA and active flushes, respectively.
More complicated schemes that fetch the data from the overflow heap are possible;
the details are outside the scope of the paper.

\textsf{\textbf{Deletes }}
A simple protocol built over active puts performs deletes. 
We use a designated page $P$ marked as \texttt{W=0},
\texttt{WL}=1, \texttt{WLD}=1.
The delete implementation issues an active put. This put is targeted at $P$ and it contains a key of the element(s) to be deleted.
The IOMMU moves the keys to 
a designated access log and a specified handler uses them to remove the elements from the local volume. 

\subsection{Collecting Statistics on Memory Accesses}
\label{sec:gatherStats}

\goal{+ Motivate the design of a tool for gathering statistics}

Automatized and efficient systems for gathering various statistics are an
important research target. 
Recent work~\cite{Geambasu:2010:CAD:1924943.1924966} presents an
\emph{active} key-value store ``Comet'', where automated gathering of statistics is one
of the key functionalities. Such systems are usually implemented in the
application layer, which significantly limits their performance. Architectures
based on traditional RMA suffer from issues similar to the ones described in~\cref{sec:distHT}.

\goal{+ Say how we handle gathering of stats}

AA enables hardware-based gathering statistics. For example, to
count the number of {puts} or {gets} to a data structure, one
has to appropriately set the control bits in the PTEs that point to the
memory region where this structure is located:
\texttt{W=1}, \texttt{WL=1}, \texttt{WLD=0} (for {puts}), and
\texttt{R=1}, \texttt{RL=1}, \texttt{RLD=0} (for {gets)}. Thus, the IOMMU
ignores the data and logs only metadata
that is later processed in a handler to generate statistics; the processing
can be enforced with active flushes. 
This mechanism would also improve the performance of cache eviction schemes in memcached applications.

AA enables gathering separate statistics for each page of data.
Yet, sometimes a finer granularity could be required to count accesses
to elements of smaller sizes. In such cases one could place the
respective elements in separate pages.

\subsection{Enabling Incremental Checkpointing}
\label{sec:FTcheckpointsIncremental}

Recent predictions about mean time between failures (MTBF) of large-scale
systems indicate failures every few
hours~\cite{Besta:2014:FTR:2600212.2600224}. Fault tolerance can be achieved
with various mechanisms. In
\emph{checkpoint/restart}~\cite{Besta:2014:FTR:2600212.2600224} all processes
synchronize and record their state to memories or disks.
Traditional checkpointing schemes record the same amount of data during every
checkpoint.
However, often only a small subset of the application state changes between two
consecutive checkpoints~\cite{vasavada2011comparing}. Thus, saving all the data
wastes time, energy, and bandwidth. In \emph{incremental checkpointing} only
the modified data is recorded.  A popular scheme~\cite{vasavada2011comparing}
tracks data modifications at the page granularity and uses the dirty bit (DB)
to detect if a given page requires checkpointing.
This scheme cannot be directly applied to RMA as memory accesses performed by
remote processes are not tracked by the MMU paging
hierarchy~\cite{recio2007remote}.


AA enables incremental checkpointing in RMA codes.
Bits \texttt{W=1}, \texttt{WL}=1, \texttt{WLD}=0
(set to the data that requires checkpointing)
enable tracking the modified pages.
To take a checkpoint all the processes synchronize and process the access logs
to find and record the modified data.
Our incremental checkpointing mechanism for tracking data modifications is orthogonal to the details of synchronization
and recording; one could use any available scheme~\cite{Besta:2014:FTR:2600212.2600224}.


%

Most often both remote and local
accesses modify the memory. The latter can be tracked by the MMU and any existing method (e.g., the DB 
scheme~\cite{vasavada2011comparing}) can be used. While checkpointing, every process 
parses both the access log and the MMU page table to track both types
of memory writes.

\subsection{Reducing the Overheads of Logging}
\label{sec:FTSystem}

\goal{+ Introduce and explain logging}

Another fault tolerance mechanism for RMA is uncoordinated
checkpointing combined with \emph{logging of puts and gets} where the crashed processes repeat their work 
and replay puts and gets that modified their state before the failure; these
puts and gets are logged during the application runtime~\cite{Besta:2014:FTR:2600212.2600224}.
While logging puts is simple and does not impact performance,
logging gets wastes network bandwidth because it requires transferring additional
data~\cite{Besta:2014:FTR:2600212.2600224}.
We now describe this issue and 
solve it with AA.


\goal{++ Describe message-logging and state why in RMA it's complicated}

\begin{figure}[h!]
\centering
\includegraphics[width=0.48\textwidth]{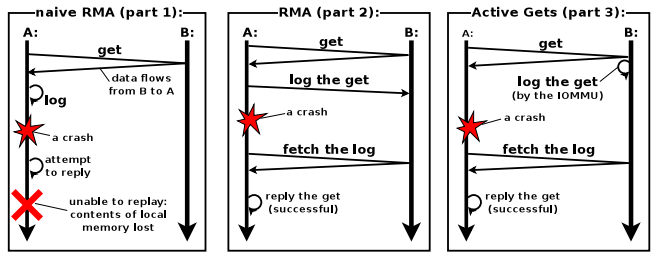}
\caption{Logging and replaying issued operations in RMA and
AA.} \label{fig:logging-gets}
\end{figure}

\textsf{\textbf{Logging Gets in Traditional RMA }}
%
%
A get issued by process A and targeted at another process B fetches data from 
the memory of B and it impacts the state of A. Thus, if A fails and begins recovery,
it has to replay this get. Still, A cannot log this get locally as the contents of its memory are lost after the crash
(see Figure~\ref{fig:logging-gets}, part~1). Thus, B can log
the get~\cite{Besta:2014:FTR:2600212.2600224}.


\goal{++ Explain why a traditional RMA logging scheme has overheads}

The core problem in RMA is that B knows nothing of {gets} issued by A, and cannot
actively perform any logging. It means that A has to wait for the data
to be fetched from B and only then can it \emph{send this data back to
B}. This naive scheme comes from the fundamental rules of one-sided RMA
communication: B is completely oblivious to any remote accesses to its memory~\cite{Besta:2014:FTR:2600212.2600224}
(cf. Fig.~\ref{fig:logging-gets}, part~2).

\begin{figure*}
\centering
 \subfloat[Performance of Netmap.]{
  \includegraphics[width=0.225\textwidth]{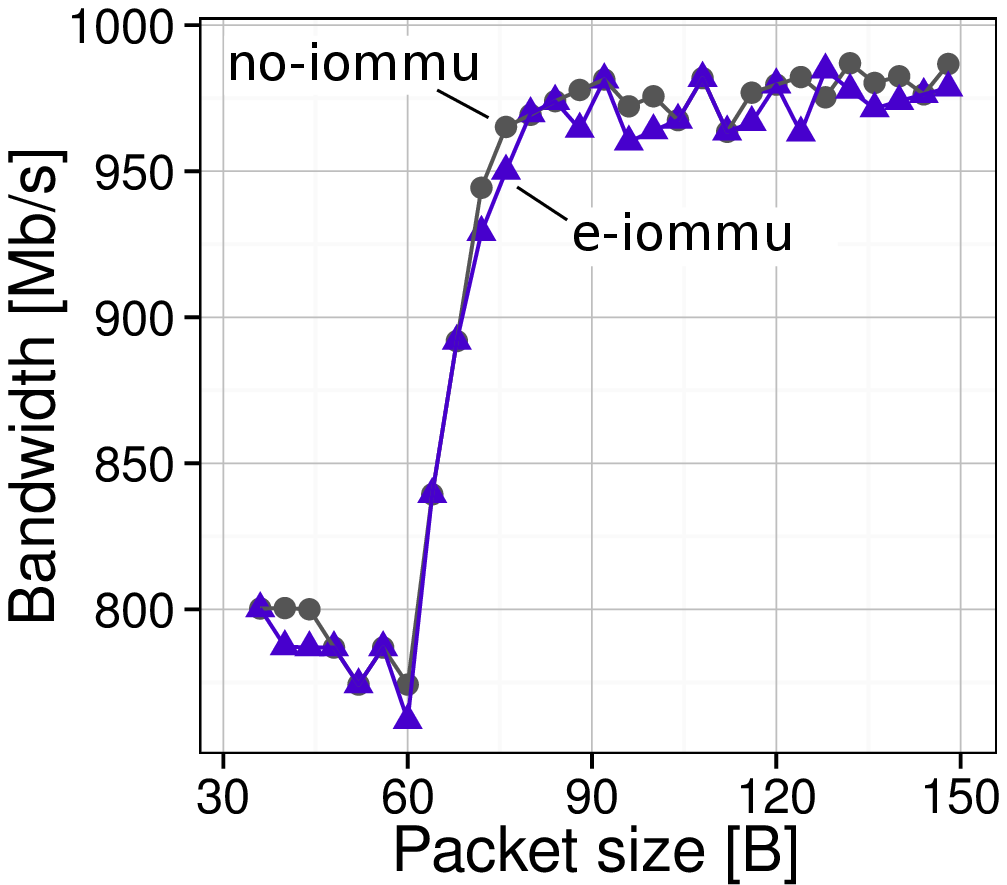}
  \label{fig:micro-netmap-tx}
 }\hfill
 \subfloat[Performance of Netperf.]{
  \includegraphics[width=0.225\textwidth]{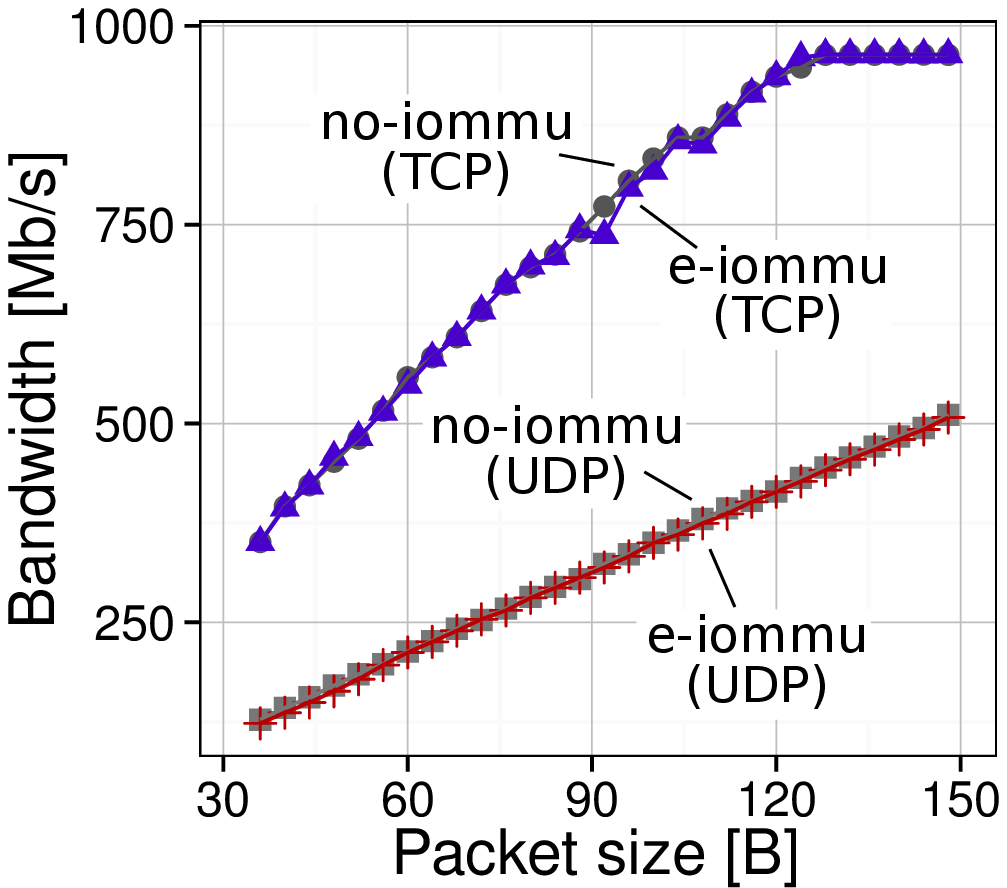}
  \label{fig:micro-netperf-tcp-udp}
 }\hfill
 \subfloat[Performance of the DHT.]{
  \includegraphics[width=0.225\textwidth]{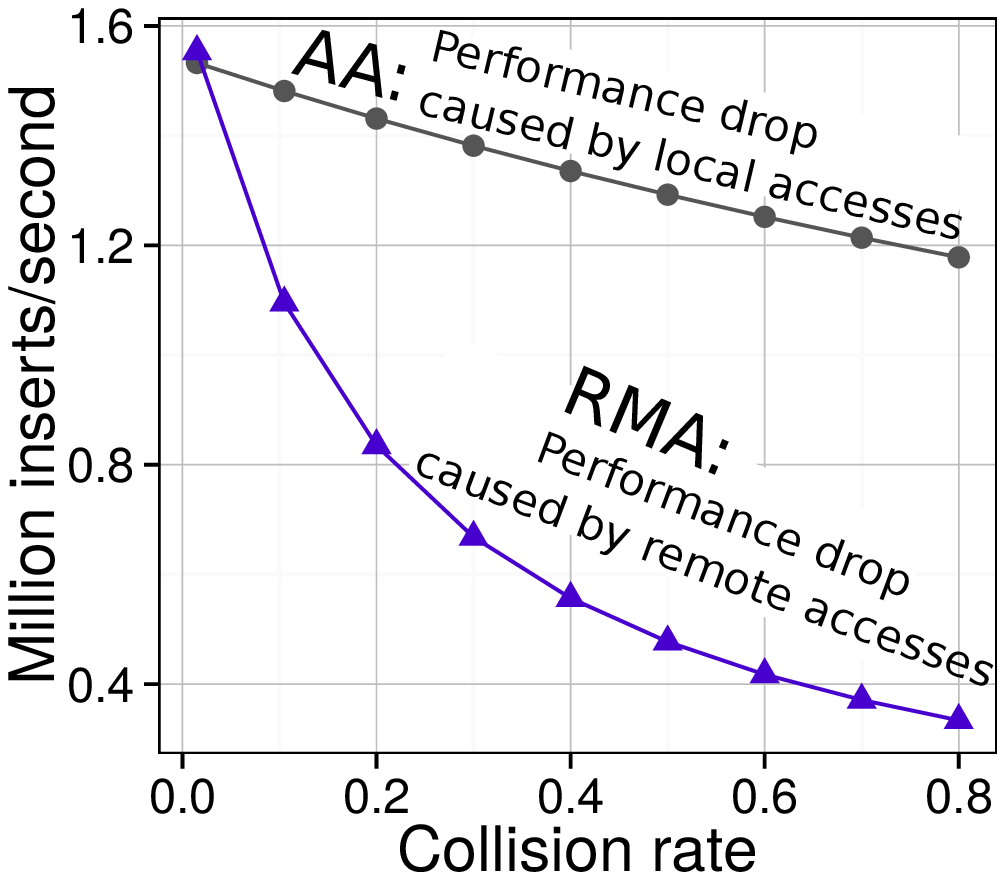}
  \label{fig:micro-dht}
 }\hfill
 \subfloat[Finding the best $C$.]{
  \includegraphics[width=0.225\textwidth]{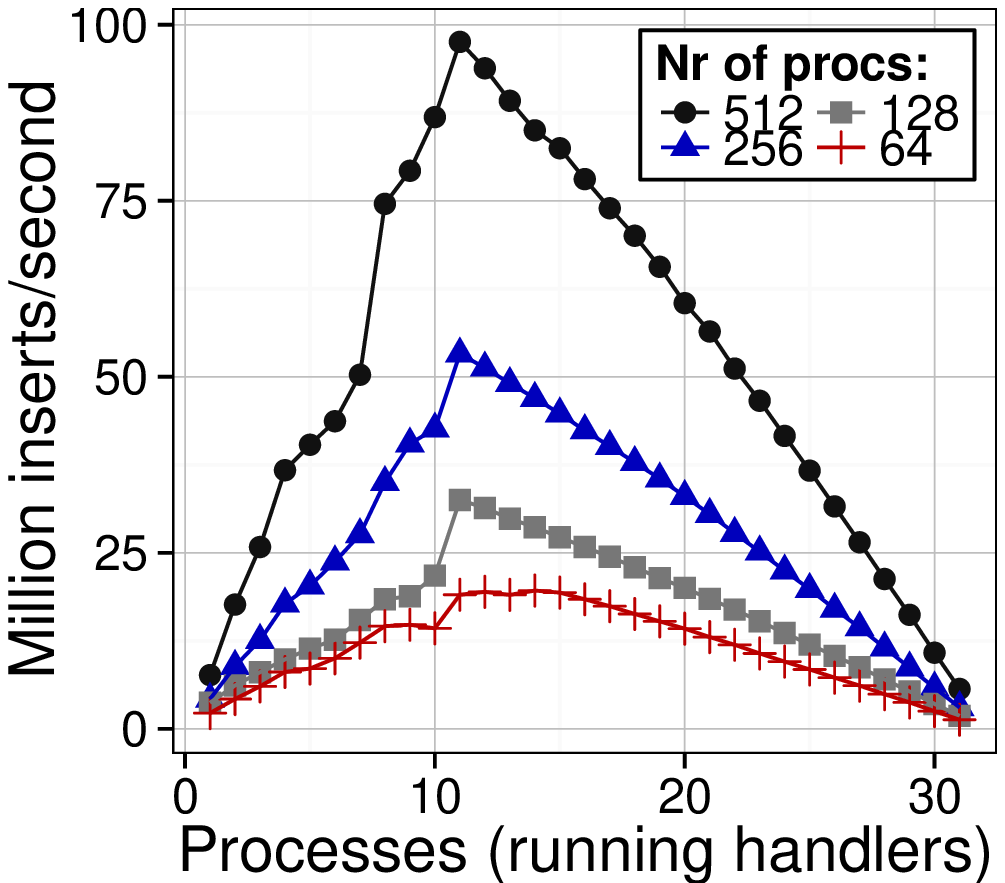}
  \label{fig:ht_inserts_cores}
 }
\caption{(\cref{sec:microbenchmarks}, \cref{sec:eval_dht}) Microbenchmarks (Figures~\ref{fig:micro-netmap-tx}-\ref{fig:micro-dht}) and finding optimum configuration for \scripttt{AA-Onload} (Figure~\ref{fig:ht_inserts_cores}).}
\label{fig:microbenchmarks}
\end{figure*}


\goal{++ Describe the logging of gets in AA}

\textsf{\textbf{Improving the Performance with Active Gets }}
In AA, the IOMMU can intercept incoming {gets}
and log the accessed data locally. First, we set up PTEs in the IOMMU page
table that point to the part of memory that is targeted by {gets}. We set
the control bits (\texttt{R=1,RL=1,RLD=1}) in these PTEs to make each {get}
touching this page {active}. Every such {get} triggers the IOMMU to copy the
accessed data into the {access log} annihilating the need for the source to
send the same data back (see Figure~\ref{fig:logging-gets}, part~3).
%
%
During the recovery a crashed process fetches the logs and then uses them to recover
to the state before the failure. We omit further details of this scheme as this
is outside the scope of this paper; example protocols (e.g., for clearing the logs or replaying puts and gets preserving the RMA consistency order) can be found in
the literature~\cite{Besta:2014:FTR:2600212.2600224}.
\section{Evaluation}
\label{sec:evaluation}

To evaluate AA we first conduct cycle-accurate simulations
to cover the details of the interaction between the NIC, the IOMMU,
the CPU, and the memory system. Second, we perform simplified large-scale
simulations to illustrate how AA impacts the performance of large-scale codes.

\subsection{Microbenchmarks}
\label{sec:microbenchmarks}

%

%
%
We first perform cycle-accurate microbenchmarks that evaluate
the performance of data transfer between
two machines connected with an Ethernet link. We compare
system configurations without the IOMMU (\texttt{no-iommu}) and with the extended IOMMU
presented in this paper (\texttt{e-iommu}).
We use the gem5 cycle-accurate full-system simulator~\cite{binkert2011gem5}
and a standard testbed
that allows for modeling two networked
machines with in-order CPUs, Intel 8254x 1GbE NICs with Intel e1000 driver, a full operating system, TCP/IP stack, and PCIe buses.
The utilized OS is Ubuntu 11.04 with precompiled 3.3.0-rc3 Linux kernel
that supports 2047MB memory.
We modify the simulated system by splitting the PCIe bus
and inserting an IOMMU in between the two parts. We model the IOMMU
as a bridge with an attached PTE cache (IOTLB).
The bridge provides buffering and
a fixed delay for passing packets; we set the delay to be 70$ns$
for each additional memory access. 
We also use a 5$ns$ delay for simulating
IOMMU internal processing. We base these values on
the L1/memory latencies of the simulated system.

We first test 
data transfer with PktGen~\cite{olsson2005pktgen} (a high-speed packet generator) enhanced with netmap\cite{rizzo2012netmap}
(a framework for fast packet I/O).
Second, we evaluate the performance of a TCP and a UDP stream with netperf, a popular benchmark
for measuring network performance. We show the results in Figures~\ref{fig:micro-netmap-tx}-\ref{fig:micro-netperf-tcp-udp}. The IOMMU presence only marginally affects the data transfer bandwidth (the difference between the \texttt{no-iommu} and \texttt{e-iommu} is 1-5\% with \texttt{no-iommu}, as expected, being marginally faster).

We also simulate a hashtable workload
of one process inserting new elements at full link bandwidth into the memory of the remote machine;
see Figure~\ref{fig:micro-dht}.
Here, we compare \texttt{AA} with a traditional \texttt{RMA} design of the DHT.
As the collision rate increases, the performance of both designs drops due to a higher
number of memory accesses. Still, \texttt{AA} is $\approx$3 times more performant than \texttt{RMA}.

%

\subsection{Evaluation of Large-Scale Applications}
\label{sec:evaluation}

\goal{Describe how we simulate RMA}


The second performance-related question is how the AA semantics,
implemented using the proposed Active Access and the IOMMU design,
impacts the performance of \emph{large-scale} codes.
%
%
%
To be able to run large-scale benchmarks on a real supercomputer,
we simplify the simulation infrastructure.
We simulate one-sided RMA accesses
with MPI point-to-point messages. We replace one-directional RMA
{puts} with a single message and two-directional RMA calls ({gets} and
atomics) with a pair of messages exchanged by the source
and target, analogously to packets in hardware. We then emulate extended IOMMUs
by appropriately stalling message handlers.
As the IOMMU performs data replication and redirection bypassing the CPU,
there are four possible sources of such overheads: interrupts,
memory
accesses due to the logging,
IOMMU page table lookups, and
accesses to the scratchpads on the CPU.

\goal{Describe how we simulate respective overheads}


First, we determine the interrupt and memory access latencies on our system to be $3 \mu s$ and $\approx$70$ns$, respectively.
Second, we simulate the IOTLB and page table lookups varying several parameters (PTE size, associativity, eviction policy).
Finally, we assume that an access to the scratchpad to notify a polling hyperthread is equal
to the cost of an L3 access and we evaluate it to $\approx$15$ns$.

\goal{Describe the hardware configuration}

All the experiments are executed on the CSCS Monte Rosa Cray XE6
system. Each node contains four 8-core AMD processors (Opteron 6276
Interlagos 2.3 GHz) and is connected to a 3D-Torus Gemini network. We
use 32 processes/node and the GNU
Environment 4.1.46 for compiling.

\goal{Describe the schemes we compare to}

We compare the following communication schemes:

\begin{description}[leftmargin=0.5em]
  \itemsep=0pt
\item[\textsf{AA-Int, AA-Poll, AA-SP:}] AA based on the IOMMU communicating with the CPU using:
interrupts, polling the main memory, and accessing the scratchpad, respectively.
\item[\textsf{RMA:}] traditional RMA representing RDMA architectures.
\item[\textsf{AM:}] an AM scheme in which processes
poll at regular intervals to check for messages.
Note that this protocol is equivalent to traditional message passing.
\item[\textsf{AM-Exp:}] an AM variant based on exponential backoff to reduce polling overhead. 
If there is no
incoming message, we double the interval after which a process will poll.
\item[\textsf{AM-Onload:}] an AM scheme where several cores are only dedicated to running
AM handlers and constantly poll on flags that indicate whether new AMs have to
be processed.  
\item[\textsf{AM-Ints:}] an AM mechanism based on interrupts generated by the NIC that
  signal to the CPU it has to run the handler.
\end{description}

\begin{figure*}
\centering

 \subfloat[DHT, $R_{cols} \approx$5\%.]{
  \includegraphics[width=0.225\textwidth]{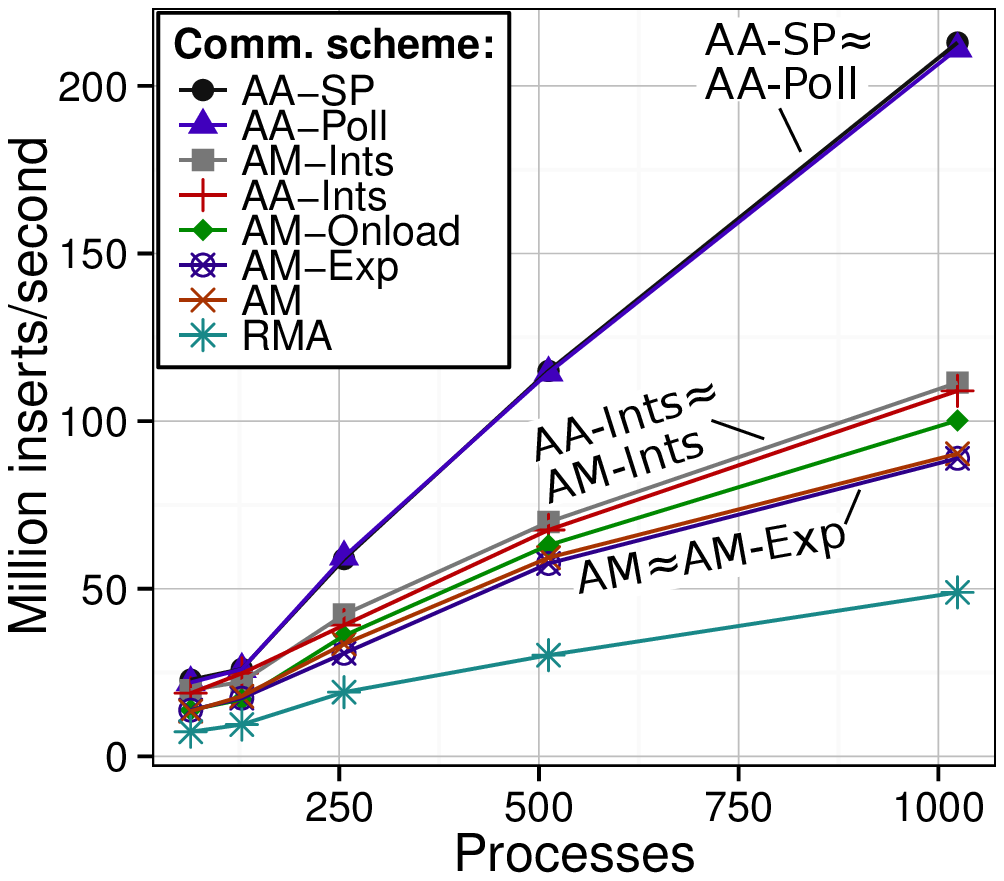}
  \label{fig:ht_inserts_tests_collisions_5}
 }\hfill
 \subfloat[DHT, $R_{cols} \approx$25\%.]{
  \includegraphics[width=0.225\textwidth]{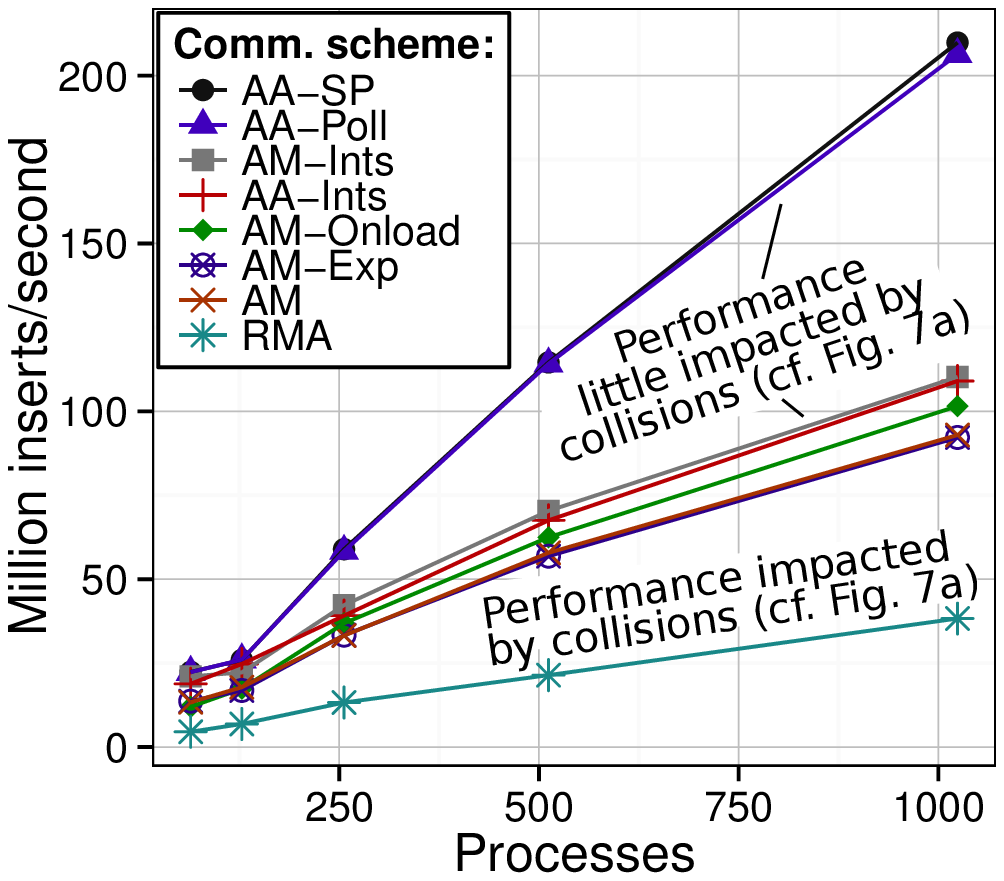}
  \label{fig:ht_inserts_tests_collisions_25}
 }\hfill
 \subfloat[DHT, IOTLB analysis.]{
  \includegraphics[width=0.225\textwidth]{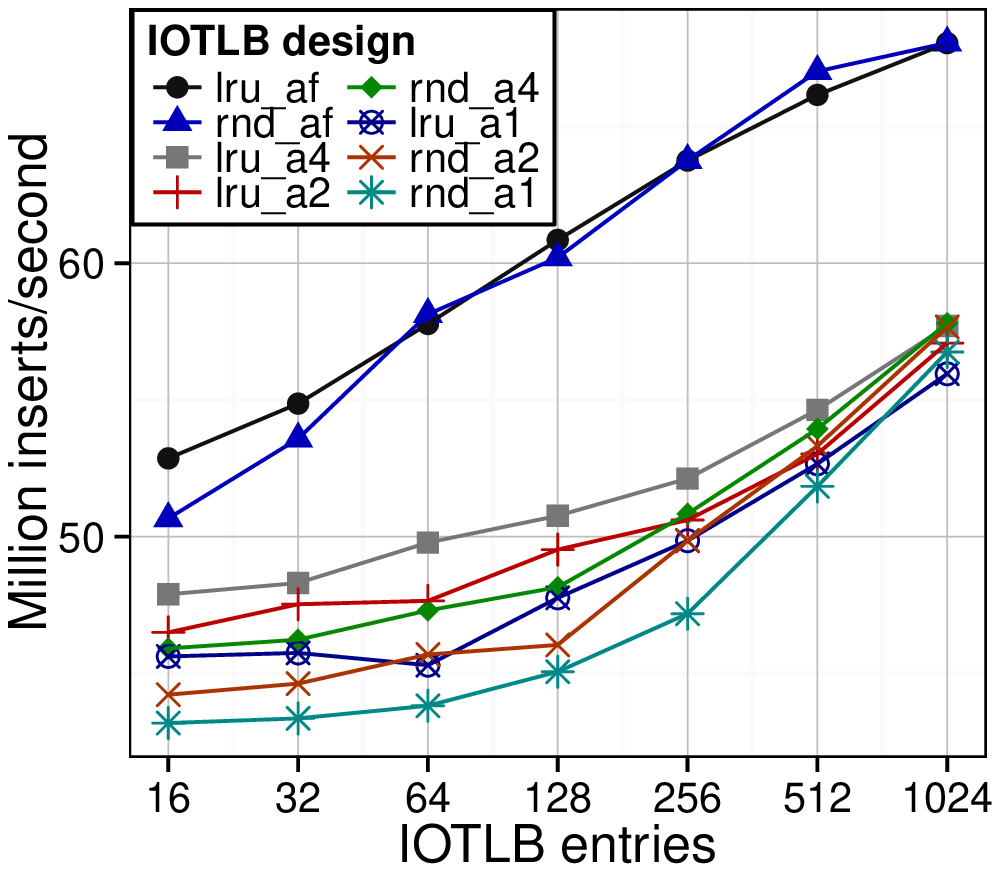}
  \label{fig:iotlb_perf}
 }\hfill
   \subfloat[Access Counter results.]{
  \includegraphics[width=0.225\textwidth]{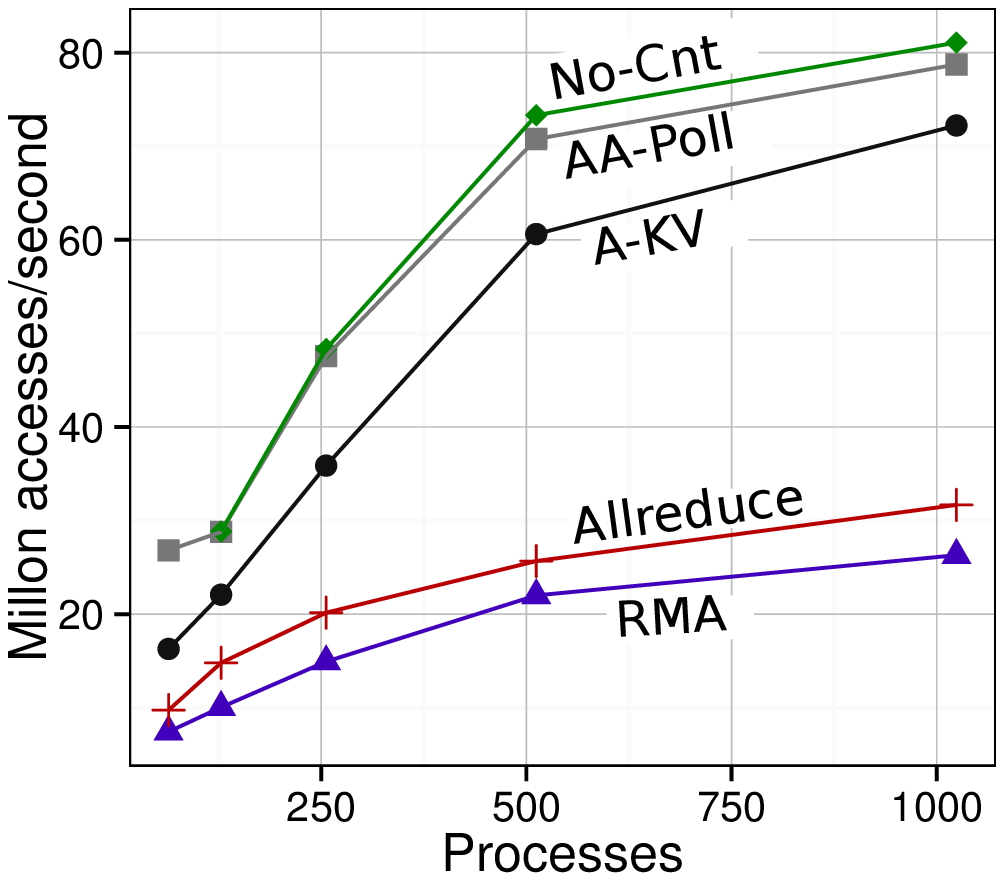}
  \label{fig:counter_perf}
 }
\caption{(\cref{sec:eval_dht}, \cref{sec:eval_ac}) The performance of the DHT (Figures~\ref{fig:ht_inserts_tests_collisions_5},~\ref{fig:ht_inserts_tests_collisions_25},~\ref{fig:iotlb_perf}) and the access counter (Figure~\ref{fig:counter_perf}). We use 32 processes/node.}
\label{fig:ht_runs_1}
\end{figure*}

\subsubsection{Distributed Hashtable}
\label{sec:eval_dht}

\goal{++ Describe the DHT benchmark}

We implement eight hashtable variants using the above schemes.
Processes insert random
elements with random keys (deletes give similar performance and are skipped). Each DHT volume can
contain $2^{21}$ elements. We vary different parameters to cover a broad
spectrum of possible scenarios. First, we study the scalability 
by changing the number of inserting processes $P$. Second, we evaluate
benchmarks with different numbers of hash collisions ($R_{cols}$, the ratio
between the number of hash collisions and the total insert count). Third, we
simulate different applications by varying computation ratios
($R_{comp}$, the ratio between the time spent on local
computation and the total experiment runtime).
We also vary the IOTLB parameters: IOTLB size,
associativity, and the eviction policy.
%
Finally, we test two variants of \texttt{AA-Ints} and \texttt{AM-Ints}
in which an interrupt is issued every $10$
and $100$ inserts (performance differences
were negligible (<5\%); we only report numbers for the former).

\texttt{AM-Onload}
depends on the number of cores ($C$) per node that are dedicated to processing AM
requests. Thus, for a fair 
comparison, we run \texttt{AM-Onload} for every $C$ between 1 and 31
to find the most advantageous configuration for every experiment. 
Figure~\ref{fig:ht_inserts_cores} shows that $C=11$ delivers maximum
performance. If $C<11$
the cores become congested and the performance decreases. $C>11$
limits performance as receiving cores become underutilized. 

\goal{+++ Describe the results when varying P and R\_\{cols\}}

\textsf{\textbf{Varying $P$ and $R_{cols}$}}
Figure~\ref{fig:ht_inserts_tests_collisions_5} shows the results for
$R_{cols}=5\%$ and Figure~\ref{fig:ht_inserts_tests_collisions_25}
for $R_{cols}=25\%$. Here, both \texttt{AA-SC} and \texttt{AA-Poll}
outperform all other schemes by $\approx$2. As expected, \texttt{AA-SC} is slightly 
($\approx$1\%) more performant than \texttt{AA-Poll}.
\texttt{AA-Ints} is comparable to \texttt{AM-Ints}; both mechanisms suffer from interrupt latency
overheads.
%
%
The reasons for performance differences in the remaining schemes are as follows: in \texttt{AM-Exp}
and \texttt{AM} the computing processes have to poll on the receive buffer and,
upon active message arrival, extract the payload. In AA this is managed by
the IOMMU and the computing processes only insert the elements
into the local hashtable. \texttt{AM-Onload} devotes fewer processes to
compute and thus, even in its best configuration, cannot outperform AA.
\texttt{RMA} issues costly atomics~\cite{schweizer2015evaluating} for every insert and 6 more remote
operations for every collision, degrading performance.

\begin{figure*}
\centering
 \subfloat[Sort Power (\cref{sec:power_cons})]{
  \includegraphics[width=0.25\textwidth]{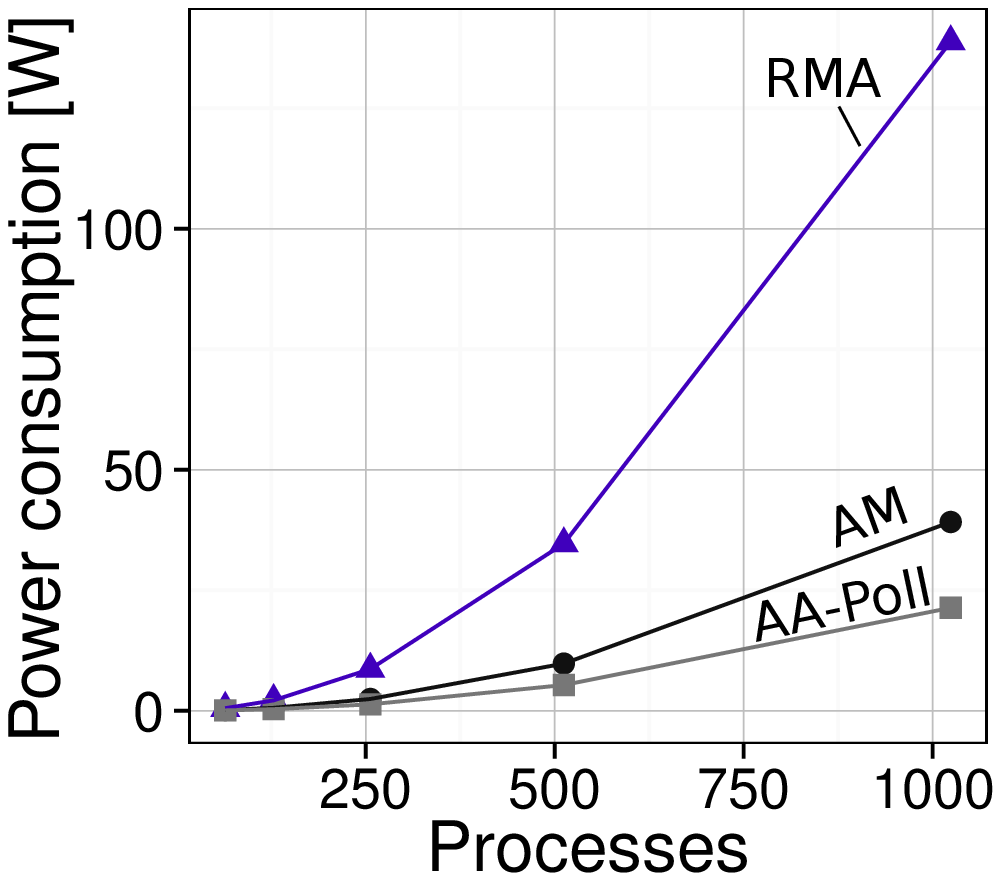}
  \label{fig:sort_pwr}
 }
 \subfloat[Power consumed by routers in parallel sort (\cref{sec:power_cons})]{
  \includegraphics[width=0.535\textwidth]{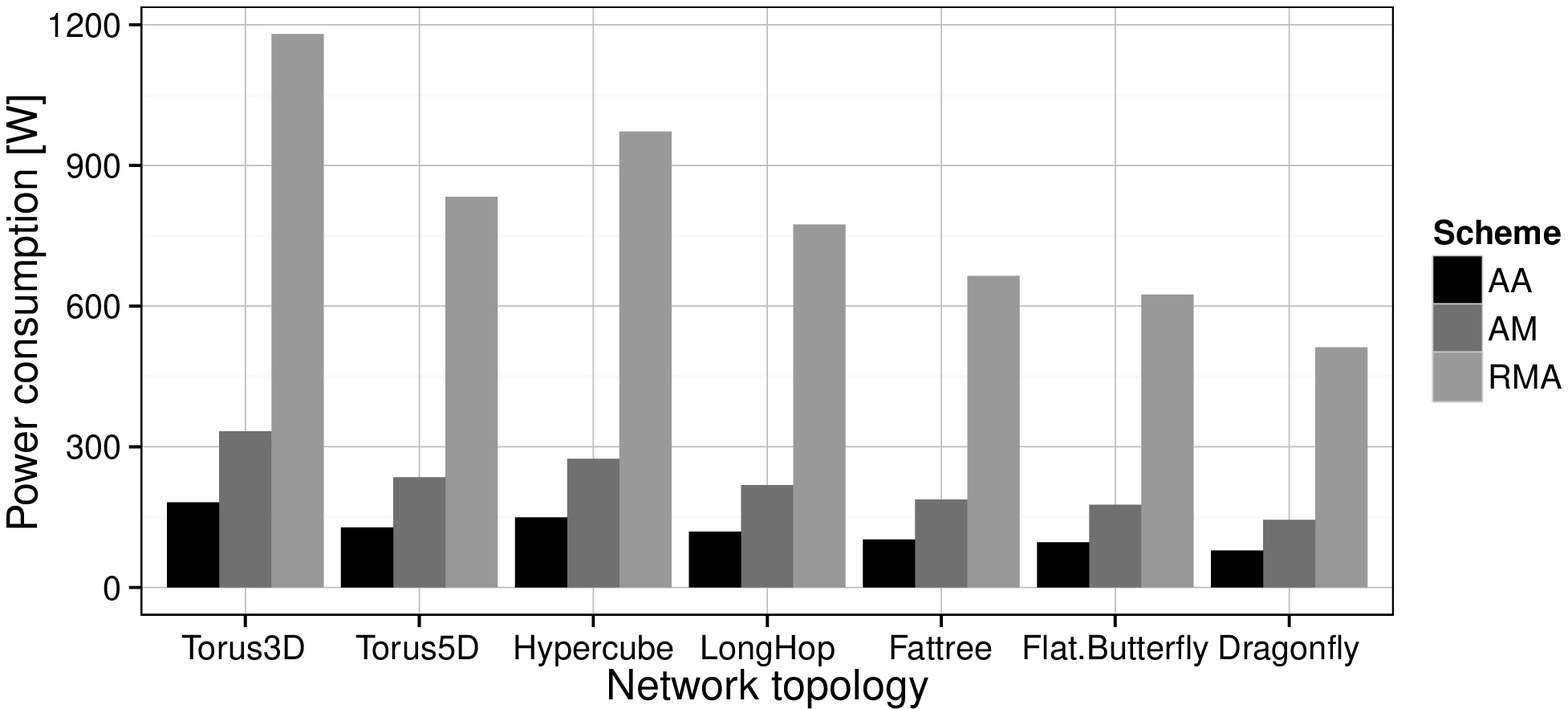}
  \label{fig:sort_pwr_topos}
 }
\caption{The power consumption advantages of the AA-based fault
tolerance scheme.}
\label{fig:ft_pwr_overall}
\end{figure*}


\goal{+++ Describe the results when varying R\_\{comp\}}

\textsf{\textbf{Varying $R_{comp}$}}
Increasing $R_{comp}$ from 0\% to 95\% did not significantly influence
the performance patterns between evaluated schemes. The only
noticeable effect was that the differences between the results of
respective schemes were smaller (which is expected as
scaling $R_{comp}$ reduces appropriately the amount of communication per
time unit).

\textsf{\textbf{Varying the IOTLB Parameters}}
We now analyze the influence of various IOTLB parameters
on the performance of DHT;
the results are presented in Figure~\ref{fig:iotlb_perf}.
The name of each plot encodes the used eviction policy (\texttt{lru}: least recently used, \texttt{rnd}: random)
and associativity  (\texttt{a1}: direct-mapped, \texttt{a2}: 2-way, \texttt{a4}: 4-way, \texttt{af}: fully-mapped).
For plot clarity we only analyze \texttt{AA-Poll}; both \texttt{AA-SP}
and \texttt{AA-Ints} follow similar performance patterns.
For a given associativity, \texttt{lru} is always better than \texttt{rnd}
as it entails fewer IOTLB misses.
Increasing associativity and IOTLB size improves the performance, for example,
using \texttt{lru\_af} instead of \texttt{lru\_a4} allows for an
up to 16\% higher insert rate.



\subsubsection{Access Counter}
\label{sec:eval_ac}

\goal{++ Describe the access counter benchmark}


We now evaluate a simple tool that counts
accesses to an arbitrary data structure. We compare \texttt{AA-Poll}
(counting done by the IOMMU), \texttt{RMA} (increasing counters with remote atomics),
and two additional designs: an approach based on the ``active
key-value'' store~\cite{Geambasu:2010:CAD:1924943.1924966} (\texttt{A-KV}), and 
a scheme where counting is done at the source and the final sums are computed with
the Allreduce collective operation~\cite{mpi3} (\texttt{Allreduce}). 
Finally, we consider 
no counting (\texttt{No-Cnt}). The number of
accesses per second is presented in Figure~\ref{fig:counter_perf}.
\texttt{AA-Poll} outperforms \texttt{A-KV} (overheads caused by
the application-level design), \texttt{RMA} (issuing costly atomics),
and \texttt{All-Reduce} (expensive synchronization).

%

\subsubsection{Performant Logging of Gets}
\label{sec:microbenchmark}


\goal{++ Describe the Simple Logging benchmark}

In the next step we evaluate the performance of {active gets} by testing
the implementation of the mechanism that logs RMA {gets}. 
Here, processes issue remote {gets} targeted at random
processes. Every {get} transfers one 8-byte integer value.
In this benchmark we do
not compare to \texttt{AM-Exp}, \texttt{AM-Onload}, and \texttt{AM-Ints} because
these schemes were not suitable for implementing this type of application.
Instead, we compare to \texttt{No-FT}: a variant with no logging (no
fault-tolerance overhead) that constitutes the best-case baseline. 
%
%
We illustrate the scalability of AA in Figure~\ref{fig:ft_micro}.
\texttt{AA} achieves the best performance, close to \texttt{No-FT}.
 In all the remaining protocols the
data to be logged has to be transferred back to a remote storage using a
{put} (\texttt{RMA}) or a {send} (\texttt{AM}), which
incurs significant overheads.
Varying the remaining parameters ($R_{comp}$, IOTLB parameters) follows the same
performance pattern as in the hashtable evaluation.


\subsubsection{Fault-Tolerant Performant Sort}
\label{sec:sort_p}

\goal{++ Describe the Parallel Sort benchmark and the results when varying P}

\begin{figure}[h!]
\centering
\subfloat[Logging gets (\cref{sec:microbenchmark})]{
  \includegraphics[width=0.22\textwidth]{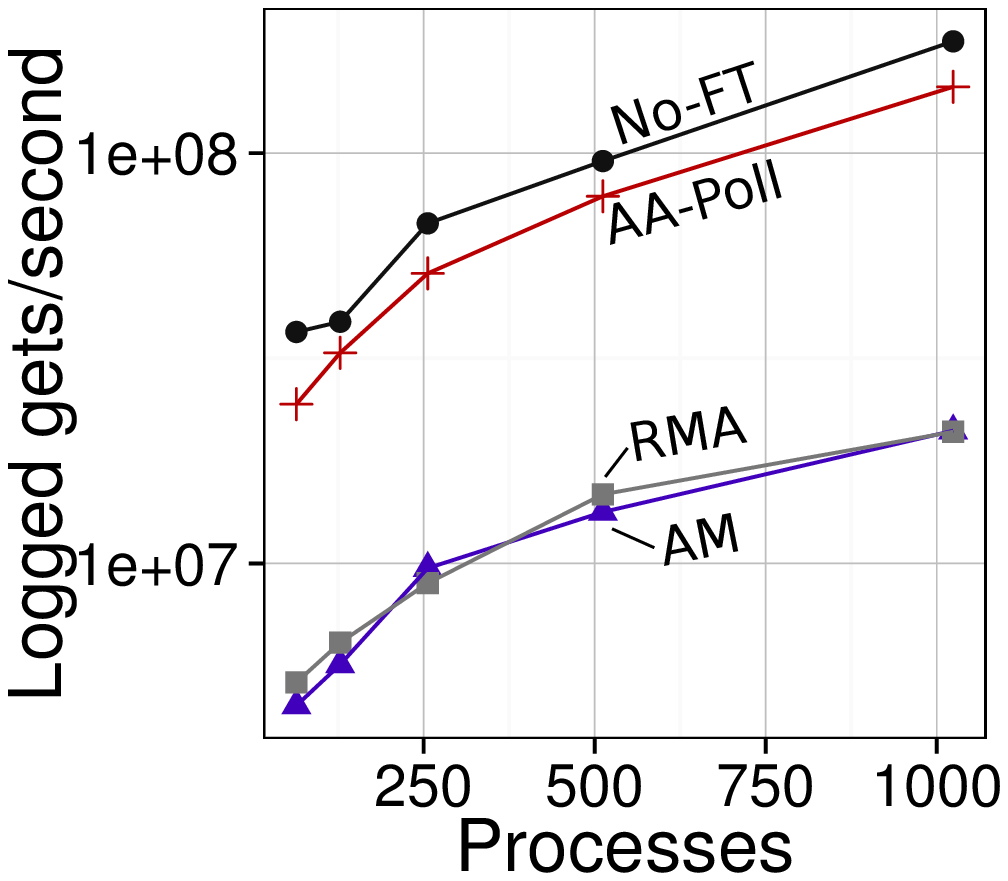}
  \label{fig:ft_micro}
 }\hfill
 \subfloat[Sort Time (\cref{sec:sort_p})]{
  \includegraphics[width=0.22\textwidth]{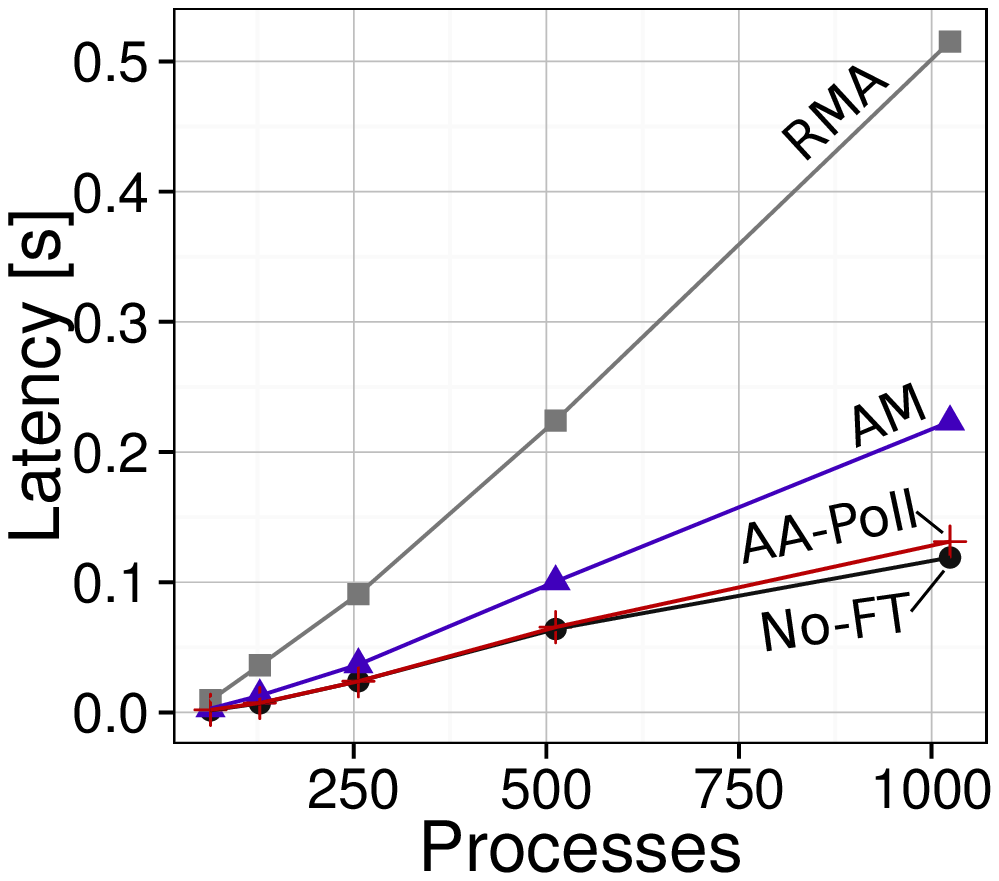}
  \label{fig:sort_lat}
 }
\caption{The performance of the AA-based fault
tolerance scheme.} 
\label{fig:ft_runs}
\end{figure}

To evaluate the performance of {active gets} we also implemented an
RMA-based version of the parallel sort Coral Benchmark~\cite{coral} that
utilizes {gets} instead of {messages}, and made it fault-tolerant. We
present the total time required to communicate the results of sorting 1GB of
data between processes in Figure~\ref{fig:sort_lat}.
Again, \texttt{AA} is close to \texttt{No-FT} ($<
1\%$) and reduces communication time by $\approx$50\% and $\approx$80\% in
comparison to \texttt{AM} and \texttt{RMA}, respectively.

\subsubsection{Lowering Power Consumption}
\label{sec:power_cons}


\goal{++ Motivate reducing power consumption}

Reducing the amount of power consumed by datacenters and HPC centers is one of
the major goals of the networking and hardware
communities~\cite{Abts:2010:EPD:1815961.1816004}. Interconnection networks are
a key factor here as their energy consumption can grow up to 50\% of the
overall energy usage of a datacenter~\cite{Abts:2010:EPD:1815961.1816004}.

\goal{++ Show and explain why we consume less energy at endpoints}

As active gets reduce the amount of communicated bytes, it also lowers
dynamic power consumption by NICs. To evaluate this effect we use the Parallel Sort
application from \cref{sec:sort_p}. We first present
(Figure~\ref{fig:sort_pwr}) the power consumed by endpoint NICs while sorting
1GB of data (we take the parameters of a single Realtek 8111E-VL-CG NIC
from~\cite{5743052}). We now make no distinction between \texttt{No-FT} and
\texttt{AA} as both schemes transfer the same amount of data over the network.
\texttt{AA} is characterized by the lowest power consumption as a direct
consequence of transmitting the smallest number of bytes.

\goal{++ Show that we're better also for any topology}

Finally, we also approximate the power consumed by the interconnection routers.
We analyze both low-radix topologies (Torus3D, Torus5D, Hypercube) and
high-radix state-of-the-art interconnects (Dragonfly~\cite{dally08}, Flat.
Butterfly~\cite{dally07}, Long Hop~\cite{2013arXiv1301.4177T}, and Fat tree).
We again assume power numbers from~\cite{5743052}. The results are
in Figure~\ref{fig:sort_pwr_topos}. The power consumed by AA is lowest for each
topology.

\section{Discussion}
\label{sec:discussion}

\goal{Introduce the section, mention there's more in TR}

We demonstrated Active Access (AA), a mechanism that uses IOMMUs to
provide ``active'' semantics for RMA and implement high-performance one-sided
communication.
Here we briefly outline additional design issues and other potential applications of AA.
%

\subsection{Additional Design Issues} \label{sec:des_issues}

\paragraph{Multiple Devices sharing the same IOMMU}
\label{sec:conc_applications}

The IOMMU design is very flexible; a single unit can cover one or multiple I/O
devices~\cite{intel_vtd}. In the latter case, to distinguish between different
devices while performing the remapping, the IOMMU could process
the PCIe transaction layer header, which contains, among others: the PCI
\emph{bus number} and the \emph{device number}. By recording these values
together with RMA operations, the AA mechanism would
distinguish between the data coming from different I/O sources.

\paragraph{Portability of AA vs. AM} 


Active message libraries face several portability problems. DCMF, PAMI, LAPI
and MX were designed and hand-tuned only for the specific machines (e.g.,
BlueGene/P and BlueGene/Q). GASNet is more portable, but this is because it
actually provides separate implementations for every different target machine.
On the other hand, two \emph{truly} portable implementations (AM over UDP and
AM over MPI) suffer from performance penalties introduced by underlying MPI
and UDP. Thus, there is no \emph{hardware mechanism} which is the
common base for different efficient AM implementations. On the other hand, AA
is constructed upon IOMMUs, which have almost a uniform design and very
similar interfaces (for example, see~\cite{intel_vtd} and~\cite{amd_iommu}).
This suggests that AA could be easily adopted by different hardware vendors.

\subsection{Potential Applications}
\label{sec:use-cases-disc}

\paragraph{Enabling Flexible Protection of Data}
\label{sec:enablingSecurity}

IOMMUs are used to protect data
by blocking potentially harmful accesses. AA could
extend this functionality and enable the user to decide
whether to log the metadata of a blocked access to gather statistics
(by setting \texttt{W=0}, \texttt{WL=1}, \texttt{WLD=0} (for {puts}), and
\texttt{R=0}, \texttt{RL=1}, \texttt{RLD=0} (for {gets})),
or discard the metadata
to save energy and reduce logging overheads
(by setting \texttt{WL=0} and \texttt{RL=0}).

\paragraph{Distributed Hardware Transactions}
\label{sec:hdtmPointer}

Transactional Memory (TM)~\cite{Herlihy:1993:TMA:165123.165164} is a well-known
technique in which portions of code (\emph{transactions}) can only be executed
atomically, as a single operation. TM can be based on software emulation or
native hardware support. Software TM was implemented in distributed
environment~\cite{Herlihy:2005:DTM:2162318.2162344}, however no hardware-based
distributed transactional memory (D-HTM) exists. AA
may enable such a scheme. First, the semantics would have to be extended with
calls such as \texttt{start\_transaction()} and \texttt{commit\_transaction()}.
To implement the functionality, one could again extend the IOMMU
to maintain an associative data
structure for temporarily storing RMA {puts} and {gets} and mapping
them to the targeted memory regions in the network. When intercepting any
remote access, the IOMMU would process the structure and check for any
potential conflicts. If there are any, it would mark the respective {gets}
and {puts} as invalid and interrupt the CPU to notify remote processes
that their transactions failed. To execute a transaction, the CPU would
issue the operations stored in the associative data structure.

\paragraph{Enabling Performant TCP Stacks}
\label{sec:fasttcp}

Significant effort has been made to improve the performance of networking
stacks. Current approaches suffer from performance penalties
caused by software-based designs or too many protocol layers (e.g., MegaPipe~\cite{Han:2012:MNP:2387880.2387894} or mTCP~\cite{Jeong:2014:MHS:2616448.2616493}),
the necessity to modify existing NIC architectures (e.g., Arsenic~\cite{pratt2001arsenic} or
SENIC~\cite{Radhakrishnan:2014:SSN:2616448.2616492}), the requirement to use expensive
proprietary hardware (e.g., IB~\cite{IBAspec}, SDP~\cite{goldenberg2005zero},
FM~\cite{pakin1995high}, or VIA~\cite{buonadonna1998implementation})
and breaking the socket semantics (e.g., Queue Pair IP~\cite{Buonadonna:2002:QPI:545215.545243}, or
RoCE~\cite{roce})
AA enables combining the best of the presented options: the utilization of the IOMMU
ensures hardware support and thus high performance; still, no expensive proprietary
hardware has to be purchased and socket semantics are preserved. The core notion
is that the access log is a FIFO data structure that could become a container
for incoming packets. Fault tolerance could be achieved by, e.g., utilizing
Lossless Ethernet.

\paragraph{Improving the Performance of Checkpointing}
\label{sec:FTcheckpoints}

\goal{+ Introduce fault-tolerance and checkpointing}

Checkpointing is often done by computing an XOR checksum of the relevant data structure and storing it
at an additional process, similarly to what is done in RAID5~\cite{Chen:1994:RHR:176979.176981}.
In RMA, processes
update the remote checksum with atomics implemented
in the remote NIC~\cite{Besta:2014:FTR:2600212.2600224}). Each such atomic
requires at least two PCIe transactions; if $N$ processes perform a checkpoint,
this would result in $2NK$ PCIe transactions, where $K$ is the number of transactions
required to transfer the data issued by one process.
AA enables a more performant scheme in which each checkpointing process
sends the whole checkpoint data to process $C$
with a single put, targeting the checksum.
The control bits in the PTEs that point to the
memory region where this checksum is located are set to the following values:
\texttt{W=0}, \texttt{WL=1}, \texttt{WLD=1}.
The data from each process is intercepted by the IOMMU and recorded in the log.
It entails issuing at most $NK$ PCIe transactions, half of what is required in
the traditional variant.

\paragraph{Hardware Virtualization of Remote Memories}

\goal{+ Say why virtual memory is cool}

\begin{figure}[h!]
\centering
\includegraphics[width=0.48\textwidth]{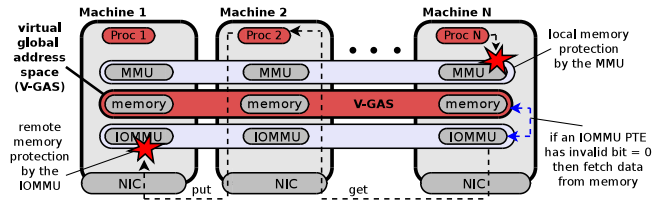}
\caption{V-GAS together with some example features.}
\label{fig:vgas}
\end{figure}

Finally, AA's potential can be further explored to provide
hardware virtualization of remote memories.
There are three major advantages of virtual memory: it
enables an OS to swap memory blocks into disk, it facilitates the application
development by providing processes with separate address
spaces, and it enables useful features such as memory protection or
dirty bits.
\goal{+ Motivate and describe virtual networked memories}
Some schemes (e.g., PGAS languages) emulate a part of these functionalities
for networked memories. 
Extending AA with features specific to MMU PTEs (e.g.,
invalid bits) would enable a hardware-based
\emph{virtual global address space} (V-GAS) with novel enhanced paging capabilities and data-centric handlers
running transparently to any code accessing the memory; see Fig.~\ref{fig:vgas}.


The IOMMUs could also become the basis of V-GAS for Ethernet. All the
described IOMMU extensions are generic and do not rely on any specific NIC
features, leaving the possibility of moving the V-GAS potential into commodity
machines that do not provide native RDMA support. For example, by utilizing
Single Root I/O Virtualization (SR/IOV), a standard support for hardware
virtualization combined with multiple receive and transmit rings, one can
utilize IOMMUs to safely divert traffic right into userspace.

\section{Related Work}
\label{sec:relatedWork}

Not all possible use cases for IOMMUs have been studied so far.  Ben-Yehuda et
al.~\cite{ben-yehuda} discuss IOMMUs for virtualization in Linux.  Other works
target efficient IOMMU emulation~\cite{Amit:2011:VEI:2002181.2002187}, reducing
IOTLB miss rates~\cite{Amit:2010:ISM:2185870.2185897}, isolating Linux device
drivers~\cite{Boyd-Wickizer:2010:TMD:1855840.1855849}, and mitigating IOMMUs'
overheads~\cite{ben2007price}.  There are also vendors' white papers and
specifications~\cite{amd_gart,amd_iommu,intel_vtd,pci_iommu,arm_iommu,Armstrong:2005:AVC:1148882.1148885,ibm_tce}.
Our work goes beyond these studies by proposing a new a mechanism and a
programming model that combines AM with RMA and uses IOMMUs for
high-performance distributed data-centric computations.

There are several mechanisms that extend the memory subsystem
to improve the performance of various codes. Active Pages~\cite{Oskin:1998:APC:279358.279387}
enable the memory to perform some simple operations allowing the CPU to
operate at peak computational bandwidth. Active Memory~\cite{Fang:2007:AMO:1274971.1275004}
and in-memory computing~\cite{zhu2013accelerating}
add simple compute logic to the memory controller and the memory itself, respectively.
AA differs from these schemes as it targets distributed RMA computations
and its implementation only requires minor extensions to the
commodity IOMMUs.

Scale-Out NUMA~\cite{Novakovic:2014:SN:2541940.2541965} is an architecture, programming model,
and communication protocol that offers low latency of remote memory accesses.
It differs from AA as it does not provide the active semantics
for both puts and gets and it introduces significant changes to the memory subsystem.
  %

%
Active messages were introduced by von Eicken et al~\cite{von1992active}.
Atomic Active Messages~\cite{besta2015accelerating}, a variant of AM, accelerate
different graph processing workloads~\cite{besta2017push} by running handlers atomically in response to the
incoming messages. The execution of the handlers is atomic thanks to hardware
transactional memory. While AM and AAM focus on incoming messages, AA specifically targets
RMA puts and gets, and its design based on IOMMUs is able to process single packets.
AA could also possibly be used to accelerate irregular distributed workloads
such as graph databases~\cite{besta2019demystifying}, general distributed graph
processing~\cite{besta2017push, gianinazzi2018communication,
solomonik2017scaling, lumsdaine2007challenges, malewicz2010pregel}, or deep
learning~\cite{ben2019modular}.
%
%
Scalable programming for RMA was discussed in different works~\cite{fompi-paper, schmid2016high}.
Some of AA's functionalities could be achieved using RMA/AM interfaces
such as Portals~\cite{portals4}, InfiniBand~\cite{IBAspec}, or GASNet~\cite{bonachea2002gasnet}.
However, Portals would introduce additional memory overheads per NIC because it 
requires descriptors for every memory region. These overheads may grow 
prohibitively for multiple NICs. Contrarily, AA uses a single 
centralized IOMMU with existing paging structures, ensuring no 
additional memory overheads. 
%
%
Furthermore, AA offers notifications on gets and it enables various
novel schemes such as incremental
checkpointing for RMA and performant logging of gets.
%
%

\goal{Describe work into IOMMUs}


AA could also be implemented in the NIC~\cite{hoefler2017spin, di2019network, firestone2018azure}. Still, using IOMMUs provides
several advantages. 
Modern IOMMUs are integrated with the memory controller/CPU and thus can be 
directly connected with CPU stratchpads for a high-performance notification 
mechanism (see~\cref{sec:interruptingCPU}). This way, all I/O devices
could take advantage of this functionality (e.g., Ethernet RoCE NICs).
Moreover, we envision other future mechanisms that would enable even further
integration with the CPU.
For example, the IOMMU could be directly connected to the CPU instruction pipeline to 
directly feed the CPU with handler code. 
Finally, one could implement AA using reconfigurable
architectures~\cite{gao2009hardware, besta2019graph, besta2019substream,
de2018designing, licht2018transformations}.  We leave these directions for
future work.

\section{Conclusion}
\label{sec:conclusion}

\goal{State RMA is becoming popular but it has some problems}

RMA is becoming more popular for programming datacenters and HPC computers.
However, its traditional one-sided model of communication may incur performance
penalties in several types of applications such as DHT.

To alleviate this issue we propose the Active Access scheme that
utilizes IOMMUs to provide hardware support for \emph{active}
semantics in RMA. For example, our AA-based DHT implementation offers a speedup of two
over optimized AMs. The novel AA-based fault tolerance
protocol enables performant logging of gets and adds negligible (1-5\%) overheads to the application runtime.
Furthermore, AA enables new schemes such as incremental checkpointing in RMA. 
Finally, our design bypasses the OS and enables more effective programming
of datacenters and HPC centers.

\goal{Describe broader potential behind AA programming model}

AA enables a new programming model that combines the benefits of one-sided
communication and active messages. AA is \emph{data-centric} as it
enables triggering handlers when certain data is accessed.
Thus, it could be useful for future data processing and analysis schemes and protocols.

\goal{Describe broader potential behind AA design}

\goal{Describe concrete future applications}

The proposed AA design, based on IOMMUs, shows the potential
behind currently available off-the-shelf hardware for developing novel
mechanisms. By moving the notification functionality from the NIC to the IOMMU
we adopt the existing IOMMU paging structures and we eliminate the need for expensive memory descriptors
present in, e.g., Portals, thus reducing memory overheads.
The IOMMU-based design may enable even more performant notification
mechanisms such as direct access from the IOMMU to the CPU pipeline.
Thus, AA may play an important role in designing efficient codes and
OS/runtime in large datacenters, HPC computers, and highly parallel manycore
environments which are becoming commonplace even in commodity off-the-shelf
computers.

%

{ 
\vspace{0em}\section*{Acknowledgements}
We thank the CSCS team granting access to the Monte Rosa machine, and
for their excellent technical support. We thank
Greg Bronevetsky (LLNL) for inspiring
comments and Ali Saidi for help with the gem5 simulator.
We thank Timo Schneider
for his immense help with computing infrastructure at SPCL.}

{\normalsize
\bibliographystyle{abbrv}
{\bibliography{references}}
}

\end{document}